\documentclass[aps,reprint,groupedaddress]{revtex4-2}

\usepackage{graphicx}
\usepackage{float}
\usepackage{xcolor}
\usepackage[colorlinks]{hyperref}
\hypersetup{
    colorlinks=true,
    linkcolor=blue,
    filecolor=blue,      
    urlcolor=blue,
    citecolor=blue
  }
  
\usepackage[utf8]{inputenc}
\usepackage[T1]{fontenc}

\usepackage{lmodern}

\usepackage{physics}

\usepackage{amssymb}
\usepackage{mathtools}
\renewcommand{\var}[1]{\sigma_{#1}^2}
\newcommand{\numberthis}{\addtocounter{equation}{1}\tag{\theequation}}
\newcommand{\Z}{\mathbb{Z}}
\newcommand{\conj}[1]{#1^*}
\newcommand{\hc}{\text{H.c.}}

\newcommand{\iu}{\text{i}}
\newcommand{\e}{e}

\begin{document}

\title{A complete POVM description of multi-channel quantum electro-optic sampling with monochromatic field modes}

\author{Emanuel Hubenschmid}
\email[]{emanuel.hubenschmid@uni-konstanz.de}

\author{Thiago L. M. Guedes}
\email[]{thiago.lucena@uni-konstanz.de}
\affiliation{Department of Physics, University of Konstanz, D-78457 Konstanz, Germany}

\author{Guido Burkard}
\email[]{guido.burkard@uni-konstanz.de}
\affiliation{Department of Physics, University of Konstanz, D-78457 Konstanz, Germany}

%\date{\today}

\begin{abstract}
	We propose a multi-channel version of quantum electro-optic sampling involving monochromatic field modes. It allows for multiple simultaneous measurements of arbitrarily many $\hat{X}$ and $\hat{Y}$ field-quadrature for a single quantum-state copy, while independently tuning the interaction strengths at each channel. In contrast to standard electro-optic sampling, the sampled mid-infrared (MIR) mode undergoes a nonlinear interaction with multiple near-infrared (NIR) pump beams. We present a complete positive operator-valued measure (POVM) description for quantum states in the MIR mode. The probability distribution of the electro-optic signal outcomes is shown to be related to an $s$-parametrized phase-space quasiprobability distribution of the indirectly measured MIR state, with the parameter $s$ depending solely on the quantities characterizing the nonlinear interaction. Furthermore, we show that the quasiprobability distributions for the sampled and post-measurement states are related to each other through a renormalization and a change in the parametrization. This result is then used to demonstrate that two consecutive measurements of both $\hat{X}$ and $\hat{Y}$ quadratures can outperform eight-port homodyne detection.
\end{abstract}

\maketitle

\section{Introduction}
Understanding simultaneous measurements of incompatible observables is key to differentiate quantum mechanics from classical physics. That a quantum theory must inevitably be of a statistical nature was already recognised in the early days of quantum mechanics and led to Heisenberg's uncertainty principle \cite{Heisenberg_1927,Kennard_1927,Werner_2019}. As a consequence thereof, it is not possible to prepare an ensemble with dispersion-free conjugate variables such as position and momentum. The quantum states of a system in conventional quantum mechanics thus cannot be represented as points in phase-space, unlike in classical mechanics, but require instead a description that captures this fundamental limitation. Here the question could arise whether the quantum-mechanical expectation value can be calculated as an ensemble average over a phase-space function, as in classical statistical mechanics. One of the most widely known attempts to formulate quantum mechanics in terms of phase-space distributions was made by Wigner \cite{Wigner_1932}. Although the quantum-state phase-space function proposed by Wigner gives the correct probabilities for the position and momentum as marginal distributions, it can take negative values and does not allow for an interpretation as a probability distribution; for this reason, it is often referred to as a quasiprobability distribution. However, the Wigner function is not the only phase-space distribution which, together with a properly chosen phase-space function representing the observable, gives the same expectation value as von Neumann's trace formula \cite[p. 207]{Neumann_2018}. In fact, there is a large family of distributions fulfilling this condition, as shown by Cohen~\cite{Cohen_1966}. %Von Neumann auf Deutsch Seite 168

Wigner later proved that there is no positive phase-space distribution, which is linear as a function of the density operator, with the correct quantum mechanical marginal distributions \cite{Wigner_1971}. Thus, if we want to directly sample a quasiprobability distribution and obtain the marginal distribution thereof, there will be some additional uncertainty. Arthurs and Kelly~\cite{Arthurs_Kelly_1965,Arthurs_1988} obtained the same result by applying von Neumann's indirect measurements \cite[ch. VI.3]{Neumann_2018} to a (specific) simultaneous measurement. Werner~\cite{Werner_2016} formalized this notion by distinguishing between preparation and measurement uncertainty and derived uncertainty relations for those.

Simultaneous measurements were realized for various physical systems, such as a transmon qubit in a microwave cavity \cite{HacohenGourgy_2016,Chantasri_2018}, an optical qubit \cite{Weston_2013,Puetz_2016,Dada_2019} or a single mode of light \cite{Walker_1986,Freyberger_1993}. In the case of the quantized electromagnetic field, the vector potential assumes the role of the generalized coordinate. For monochromatic field modes, the vector potential is proportional to the $\hat{X} = \frac{1}{2}(\hat{a} + \hat{a}^\dagger)$ quadrature, in which $\hat{a}$ is the bosonic annihilation operator of the field mode, and the corresponding canonical conjugate is the electric-field related $\hat{Y} = \frac{\iu}{2}(\hat{a}^\dagger - \hat{a})$ quadrature \cite[p.~94]{Vogel_Welsch},\cite[p.~17]{gerry_knight_2004}. This allows to describe a single-mode quantum state of the electromagnetic field mode using quasiprobability distributions. There are several optical methods to sample specific quasiprobability distributions, such as, e.g., the many variants of homodyne detection to measure the quadratures either seperatly \cite{Vogel_1989,Smithey_1993,Leonhardt_1994,Wallentowitz_1996,Breitenbach_1997,Luis_2015,Bohmann_2018,Tiedau_2018,Knyazev_2018,Olivares_2019} or simultaneously
\cite{Walker_1986,Freyberger_1993,Leonhardt_1993,Zucchetti_1996,Rehacek_2015}. One such possibility is to use eight-port homodyne detection, in which the sampled mode is split into two modes using a beam splitter, then the two quadratures can be measured simultaneously using a four-port homodyne detection scheme \cite{Freyberger_1993}. Another possibility to measure the field quadratures is provided by quantum electro-optic sampling (EOS) \cite{Gallot_1999,Riek_2015,Moskalenko_2015,Riek_2017,Kizmann_2019,Guedes_2019,BeneaChelmus_2019,Lindel_2020,Lindel_2021,Kizmann_2022,Onoe_2022}. EOS is an indirect measurement of low-frequency modes, usually in the mid-infrared (MIR), mediated by higher frequency modes, usually in the near-infrared (NIR). For this to happen, modes in the two frequency ranges are correlated by an interaction in a nonlinear crystal \cite{Namba_1961}\cite[ch.~11]{Boyd}. This configuration is limited to the measurement of a single quadrature at a time. However, the simultaneous measurement of two noncommuting quadratures is of great interest to many applications in quantum information technologies, like quantum metrology \cite{Steuernagel_2004,Steinlechner_2013,Du_2020,Ast_2016}, continuous-variable quantum teleportation \cite{Vaidman_1994,Braunstein_1998}, as well as continuous-variable quantum key distribution \cite{Weedbrook_2004,Lance_2005}.

In this paper, we show that a multi-channel version of (continuous-wave-driven) electro-optic sampling can be utilized to sample quasiprobability distributions of a monochromatic MIR quantum state and thus overcome the limitation of EOS to measurements of a single quadrature. We explicitly demonstrate this by calculating the count-probability distribution, Eq.~\eqref{eq:prob_dist}. Differently from the standard approach to electro-optic sampling, multiple monochromatic NIR modes assigned to different channels are used to probe a single MIR mode. This allows one to tune the interaction strength between the MIR and each NIR mode individually. Thus, the model presented in this paper is applicable to arbitrary many measurements of $\hat{X}$- and/or $\hat{Y}$-quadratures. Furthermore, we derive the post-measurement quasiprobability distributions for arbitrary combinations of $\hat{X}$- and/or $\hat{Y}$-quadrature measurements, Eq.~\eqref{eq:post_msnt_qpd}, and use this result to show that additional measurements on the post-measurement state can outperform eight-port homodyne detection.

In section~\ref{s:model}, the proposed measurement scheme and the respective theoretical model are presented. Then, in section~\ref{s:count_prob}, the count-probability distribution is derived and some special cases are discussed. In section~\ref{s:post}, we show how the quasiprobability distribution for the post-measurement state relates to the initial states one. Finally, in section~\ref{s:comparison}, the different measurement schemes based on electro-optic sampling are compared and it is shown that two consecutive measurements of the same state can outperform eight-port homodyne detection.

\section{Model}\label{s:model}
\begin{figure}[t]
	\centering
	\includegraphics{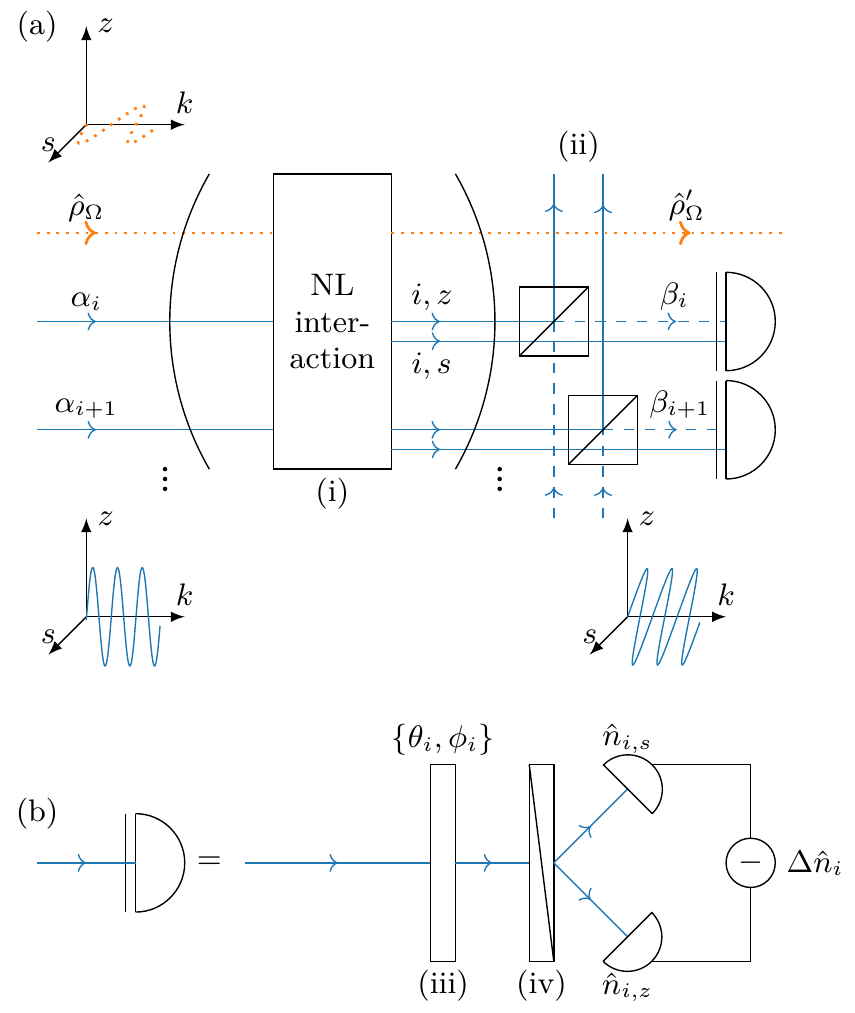}
	\caption{A schematic representation of the proposed measurement setup. The first step in (a) is a nonlinear interaction between the NIR pump beams with amplitudes $\alpha_i$ and the MIR mode $\hat{\rho}_\Omega$ in the optical parametric oscillator (i). Then, the $z$-polarized components of the NIR cavity frequencies are filtered out and replaced by $z$-polarized probe beams $\beta_i$ using the polarizing beam splitter (ii). Subsequently, the quadratures of the NIR modes are measured using the ellipsometry scheme depicted in (b). The ellipsometer consists of the $\phi_i$-wave plate (iii) rotated by $\theta_i$, a polarizing beam splitter (iv) to separate the $s$- and $z$-polarized photons of the NIR frequencies and two photon detectors to count the photon-number of each polarization. The difference between the $s$- and $z$-polarized photon-numbers $\Delta \hat{n}_i = \hat{n}_{i,s} - \hat{n}_{i,z}$ constitutes the signal.}
	\label{fig:setup}
\end{figure}

As is the case for any quantum-mechanical indirect measurement, electro-optic sampling makes use of an ancillary system (in the present case, the high-frequency NIR modes of the electric field), which becomes correlated with the low-frequency mode of the field (here the MIR) through interactions in a nonlinear crystal \cite{Namba_1961,Gallot_1999,Riek_2015,Moskalenko_2015,Riek_2017,Kizmann_2019,Guedes_2019,BeneaChelmus_2019,Lindel_2020,Lindel_2021,Kizmann_2022,Onoe_2022}. We consider the specific case of an optical parametric oscillator consisting of a zincblende-type nonlinear crystal in a cavity, labelled as (i) in Fig.~\ref{fig:setup} (a).
The details about the geometrical arrangement of the nonlinear crystal are given in \cite{Moskalenko_2015}. In the crystal, the coherent, $z$-polarized pump beams in the NIR cavity modes $i \in I=\{1,2,\ldots\}$, with amplitudes $\alpha_i$, drive the entanglement between the $s$-polarized NIR modes, corresponding to the bosonic operator $\hat{a}_{i,s}$, and the $s$-polarized MIR mode, related to $\hat{a}_{\Omega,s}$. This nonlinear interaction is described by the multi-mode squeezing operator \cite[p.276-281]{Vogel_Welsch}
\begin{equation}\label{eq:two_mode_squeezing}
	\hat{U}_\text{NL}= \exp(\conj{\zeta}\hat{a}_{\Omega,s}\sum_{i \in I} \tilde{\alpha}_i \hat{a}_{i,s}-\hc),
\end{equation}
where $\zeta$ is the dressed squeezing parameter and $\tilde{\alpha}_i = \alpha_i/(\sum_{i \in I} \abs{\alpha_i}^2)^{1/2}$ the normalized pump amplitudes. The time-evolution operator~\eqref{eq:two_mode_squeezing} is therefore an effective two-mode squeezing operator between $\hat{a}_{\Omega,s}$ and $\sum_{i \in I} \tilde{\alpha}_i \hat{a}_{i,s}$. This effective two-mode description is achieved by absorbing the normalization constant of the $\alpha_i$ into the undressed squeezing parameter~$\zeta_0$.

After the nonlinear interaction, the $z$-polarized NIR pump beams are filtered out by the polarizing beam splitters (ii) and coherent NIR probes of amplitudes $\beta_{i}$ are introduced as replacements. The $z$-polarized NIR modes  are therefore displaced by $\hat{D}_{i,z}(\beta_{i}) = \exp(\beta_{i}\hat{a}_{i,z}^\dagger-\hc)$. This allows for an additional, tunable set of parameter. If a setup without this filtering is considered, the pump amplitudes $\alpha_i$ can be set equal to the probe amplitudes~$\beta_i$.

The quadratures of the ($s$-polarized) NIR field modes are then measured using ellipsometers \cite{Gallot_1999,Moskalenko_2015,Sulzer_2020}. The first step of the ellipsometry scheme, (iii) in Fig.~\ref{fig:setup} (b), is the change in the ellipticities of the joint-polarization NIR modes due to a $\phi_i$-wave plate rotated by an angle $\theta_i$ relative to the $z$-axis, $\hat{U}_{i,\text{WP}} = \exp(\iu\phi_i \hat{a}_{i,\theta_i}^\dagger \hat{a}_{i,\theta_i})$. $\hat{a}_{i,\theta_i} = \cos(\theta_i)\hat{a}_{i,s} + \sin(\theta_i)\hat{a}_{i,z}$ are the annihilation operators of the modes the wave plate acts on. The total time-evolution operator is thus
\begin{equation}\label{eq:time_evolution}
	\hat{U} = \hat{U}_{\text{WP}}\hat{D}_z(\vec{\beta})\hat{U}_\text{NL}
.\end{equation}
To allow for a compact notation we have introduced the total wave plate operator $\hat{U}_{\text{WP}} = \bigotimes_{i \in I} \hat{U}_{i,\text{WP}}$ and the total displacement operator $\hat{D}_z(\vec{\beta}) = \bigotimes_{i \in I} \hat{D}_{i,z}(\beta_{i})$. 

In a second step, denoted (iv) in Fig.~\ref{fig:setup} (b), the photons at each NIR cavity frequency $i$ are split spatially into $s$-polarized and $z$-polarized contributions with the aid of a polarizing beam splitter. Lastly, the photons of each polarisation are counted using photon detectors and the number of $z$-polarized counts is subtracted from the $s$-polarized ones. The corresponding observables are thus the difference between the respective photon-number operators at each cavity frequency $i$,
\begin{equation}\label{eq:observable}
	\Delta \hat{n}_i = \hat{n}_{i,s}-\hat{n}_{i,z} = \sum_{\Delta n_i=-\infty}^\infty \Delta n_i \hat{P}_{\Delta n_i}
.\end{equation}
In order to obtain a description in terms of positive operator-valued measures (POVMs), the observables are decomposed into the projectors on the subspace of states with photon-number difference $\Delta n_i$ between the $s$-polarized and the $z$-polarized NIR counts at frequency $i$,
\begin{equation}\label{eq:projectors}
	\hat{P}_{\Delta n_i} = \sum_{n_i=\tilde{n}_i}^\infty\ket{n_i+\Delta n_i}_{i,s}\prescript{}{i,s}{\bra{n_i+\Delta n_i}} \otimes \ket{n_i}_{i,z}\prescript{}{i,z}{\bra{n_i}}
,\end{equation}
where the summations starts at $\tilde{n}_i=\max\{0,-\Delta n_i\}$ to avoid negative photon-number occupations. These are the necessary components to calculate the count probability of the photon-number differences $\{\Delta n_i\} = \{\Delta n_i \mid i \in I\}$.

\section{Count-probability distribution}\label{s:count_prob}
The probability distribution
\begin{equation}\label{eq:def_prob_dist}
	p(\{\Delta n_i\}) = \tr(\hat{P}_{\{\Delta n_i\}}\hat{U}\hat{\rho}_\Omega \otimes \ket{0}_\omega \prescript{}{\omega}{\bra{0}}\hat{U}^\dagger)
\end{equation}
to measure the set of photon-number differences $\{\Delta n_i\}$ is the expectation value of the projector $\hat{P}_{\{\Delta n_i\}} = \bigotimes_{i \in I}\hat{P}_{\Delta n_i}$ with respect to the time-evolved state $\hat{U}\hat{\rho}_\Omega \otimes \ket{0}_\omega \prescript{}{\omega}{\bra{0}}\hat{U}^\dagger$. The density operator $\hat{\rho}_\Omega$ corresponds to the initial state of the $s$-polarized MIR mode of the electromagnetic field and $\ket{0}_\omega  = \bigotimes_{i \in I}\ket{0}_{i,s}\otimes\ket{0}_{i,z}$ is the vacuum of the NIR modes. The density operator is expressed in the coherent-state basis $\ket{z}_{\Omega} = \hat{D}_{\Omega}(z)\ket{0}_{\Omega}$ of the MIR-Hilbert space [which is generated by displacing the vacuum with $\hat{D}_{\Omega}(z) = \exp (z \hat{a}_{\Omega,s}^\dagger + \hc)$] using the Glauber-Sudarshan quasiprobability distribution $\rho(z;s=1)$ of the sampled MIR state \cite{Glauber_1963,Sudarshan_1963},
\begin{equation}\label{eq:thz_glauber}
	\hat{\rho}_\Omega = \int \rho(z;1) \ket{z}_\Omega\prescript{}{\Omega}{\bra{z}} \dd^2 z
.\end{equation}
The same is done with the projector $\hat{P}_{\Delta n_i}$, defined in Eq.~\eqref{eq:projectors}. The probability distribution in Eq.~\eqref{eq:def_prob_dist} is therefore expressed completely in the coherent-state basis, resulting in convolutions of the Glauber-Sudarshan distributions with the matrix elements of the evolution operator in the coherent-state basis (cf. Appendix~\ref{a:count_prob}). Furthermore, it can be shown that this is equivalent to a convolution of the MIR Glauber-Sudarshan distribution and the Skellam distribution, which is the probability distribution of the difference $\Delta n_i$ between two Poissonian counting events. This is the statistics expected from the ellipsometry scheme in Fig.~\ref{fig:setup} (b), since the signal is the difference between two (predominantly) Poissonian photon-number counting events (because the modes at the photon detectors are dominated by the coherent-probe amplitudes $\beta_{i}$ and the photon-number distribution for coherent states is Poissonian). 

We want the electro-optic signal to be balanced, which means that the signal is on average zero as long as the MIR mode is in a state for which all quadrature expectation values vanish (e.g., the vacuum). This way, noise affecting both polarizations of the NIR frequencies cancel out in the ellipsometry scheme. We achieve this by choosing the rotation angle of the $\phi_i$-wave plate as
\begin{equation}\label{eq:eos_condition}
    \theta_i = (-1)^{k_1} \frac{1}{2} \arccos\left\{(-1)^{k_2}\sqrt{\frac{1}{2}\left[1-\cot^2\left(\frac{\phi_i}{2}\right)\right]}\right\}
,\end{equation}
which has solutions for $\frac{\pi}{2} \leq \phi_i \leq \frac{3}{2}\pi$, $k_1, k_2 \in \Z$. With this choice, the complex phase determining the interference at the $i$-th ellipsometry is given by
\begin{align*}\label{eq:phi}
	&\varphi_i = (k_1+k_2+1)\pi + (-1)^{k_2}\arcsin[\sqrt{2}\cos(\phi_i/2)] \\
	&+ \arg(\zeta) - \arg(\tilde{\alpha}_i) - \arg(\beta_{i}) \mod 2\pi \numberthis
,\end{align*}
as can be seen from Eqs.~\eqref{eq:phase_1}, \eqref{eq:phase_2}, and \eqref{eq:phase_3}. Choosing the signal to be balanced additionally allows to approximate the Skellam distribution by a Gaussian. This enables an analytical solution of the convolution (see Appendix~\ref{ap:skellam} for details). 

Before the count-probability distribution is presented, we have to introduce a special family of quasiprobability distributions. The standard $s$-parametrized quasiprobability distribution $\rho(z;s) = 1/\pi^2 \int \exp(z\conj{\gamma} - \conj{z}\gamma) \chi(\gamma;s) \dd^2 \gamma$ with the $s$-parametrized characteristic function $\chi(\gamma;s) = \exp(s\abs{\gamma}^2)\ev{\hat{D}(\gamma)}$ \cite{Cahill_1969},\cite[p.~128]{Vogel_Welsch},\cite[p.~110]{Carmichael_1999} is not sufficient to capture all the different cases our model describes. This is why we have to introduce the $(s_\text{X},s_\text{Y})$-parametrized quasiprobability distribution
\begin{equation}\label{eq:two_ordering_qpd}
	\rho(z;s_\text{X},s_\text{Y}) = \frac{1}{\pi^2}\int \e^{z\conj{\gamma} - \conj{z} \gamma} \chi(\gamma;s_\text{X},s_\text{Y}) \dd^2 \gamma
,\end{equation}
defined in terms of the $(s_\text{X},s_\text{Y})$-parametrized characteristic function
\begin{equation}\label{eq:two_ordering_char}
	\chi(\gamma;s_\text{X},s_\text{Y}) = \chi(\gamma;0)\e^{\frac{s_\text{Y}}{2}\Re^2(\gamma) + \frac{s_\text{X}}{2}\Im^2(\gamma)}
.\end{equation}
This is a special case of the generalized quasiprobability distribution using a Cohen function $f(\gamma) = \exp(s_\text{X}\Re(\gamma)^2/2 + s_\text{Y}\Im(\gamma)^2/2)$ \cite{Cohen_1966}.

For $s_\text{X},s_\text{Y} < 1$, the two-parameter quasiprobability distribution can be expressed as a Weierstrass transform of the Glauber-Sudarshan distribution:
\begin{align*}\label{eq:two_ordering_qpd_glauber}
	&\rho(z;s_\text{X},s_\text{Y}) = \frac{1}{\pi}\frac{2}{\sqrt{(1-s_\text{X})(1-s_\text{Y})}} \int \rho(y;1) \\
	&\times \e^{-\frac{2}{1-s_\text{X}}\Re^2(y - z) - \frac{2}{1-s_\text{Y}}\Im^2(y - z)} \dd^2y \numberthis
.\end{align*}
We can cast the count-probability distribution after application of the Gaussian approximation to the Skellam distribution [cf. Eq.~\eqref{eq:skellam_approx}] into the form of the $(s_\text{X},s_\text{Y})$-parametrized quasiprobability distribution in Eq.~\eqref{eq:two_ordering_qpd_glauber} if we partition the set of all NIR frequencies into two disjoint sets $I_\text{X}$ and $I_\text{Y}$, with $I = I_\text{X} \bigcup I_\text{Y}$ and $I_\text{X} \bigcap I_\text{Y} = \emptyset$. If the angles of the wave plates are then chosen such that $\e^{\iu\varphi_{i_\text{X}}} = 1$ for all $i_\text{X} \in I_\text{X}$ and $\e^{\iu\varphi_{i_\text{Y}}} = \iu$ for all $i_\text{Y} \in I_\text{Y}$, then the count-probability distribution
\begin{equation}\label{eq:prob_dist}
	p(\{\Delta n_i\}) \approx N(\{\Delta n_i\})\rho(z(\{\Delta n_i\});\tilde{s}_\text{X},\tilde{s}_\text{Y})
\end{equation}
is given in terms of the two-parameter quasiprobability distribution $\rho(z(\{\Delta n_i\});\tilde{s}_\text{X},\tilde{s}_\text{Y})$, as well as a renormalization envelope 
\begin{align*}\label{eq:renomrm_prob_dist}
	&N(\{\Delta n_i\}) = \pi \frac{\sqrt{(1-\tilde{s}_\text{X})(1-\tilde{s}_\text{Y})}}{2 \sqrt{1+A_X}\sqrt{1+A_Y}} \prod_{i \in I} \frac{\e^{-\Delta n_i^2/(2\abs{\beta_{i}}^2)}}{\sqrt{2\pi \abs{\beta_{i}}^2}}  \\
	&\times \exp\left\{-\frac{2\Re^2\left[z(\{\Delta n_i\})\right]}{1+\tilde{s}_\text{X}} -\frac{2\Im^2\left[z(\{\Delta n_i\})\right]}{1+\tilde{s}_\text{Y}}\right\} \numberthis
,\end{align*}
%
%with $b = \frac{\abs{\nu}^2}{\mu^2}(\sum_{i_\text{X}}\abs{\tilde{\alpha}_{i_\text{X}}}^2 - \sum_{i_\text{Y}}\abs{\tilde{\alpha}_{i_\text{Y}}}^2)$.
with $\mu = \cosh(\abs{\zeta})$ and $\nu = \exp[\iu\arg(\zeta)]\sinh(\abs{\zeta})$ and the rescaled, combined pump strength $A_Q = 2\abs{\nu}^2\sum_{i_Q}\abs{\tilde{\alpha}_{i_Q}}^2$ where $Q$ is from now on used as a placeholder for both $X$ and $Y$. As will become clear later, the set $I_\text{X}$ ($I_\text{Y}$) corresponds to $\hat{X}$-($\hat{Y}$-)quadrature measurements.
\begin{figure}[t]
	\centering
	\includegraphics{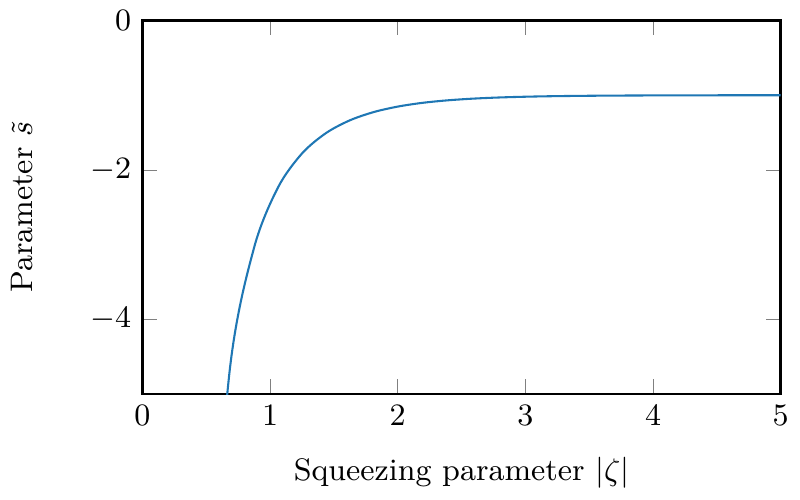}
	\caption{The parameter $\tilde{s}$ for a measurement with equal probe amplitudes, $A_X = A_Y$, as a function of the squeezing strength $\abs{\zeta}$.}
	\label{fig:order_param_squeezing}
\end{figure}
\noindent
The parameters
\begin{equation} 
	\tilde{s}_{\text{Q}} = 1-\frac{2}{1-1/\left(1+A_Q\right)}
,\end{equation}
with $Q = X,Y$, solely depend on quantities characterizing the interaction between the NIR and MIR modes. If the sum of the probe amplitudes for the $\hat{X}$ measurements equals that for the $\hat{Y}$ measurements, $A_X = A_Y = \abs{\nu}^2$, the parameter $\tilde{s} = \tilde{s}_\text{X} = \tilde{s}_\text{Y}$ is a function of the squeezing parameter $\abs{\zeta}$, as depicted in Fig.~\ref{fig:order_param_squeezing}. The argument of the quasiprobability distribution in Eq.~\eqref{eq:prob_dist} is related to the photon-number differences $\{\Delta n_i\}$ via
\begin{align*}\label{eq:z}
	&z(\{\Delta n_i\}) = -\abs{\nu}\Bigg(\frac{1+\tilde{s}_\text{X}}{2} \sum_{i_\text{X}\in I_\text{X}} \frac{\abs{\tilde{\alpha}_{i_\text{X}}}}{\abs{\beta_{i_\text{X}}}}\Delta n_{i_\text{X}} \\
	&+ \iu\frac{1+\tilde{s}_\text{Y}}{2} \sum_{i_\text{Y}\in I_\text{Y}} \frac{\abs{\tilde{\alpha}_{i_\text{Y}}}}{\abs{\beta_{i_\text{Y}}}}\Delta n_{i_\text{Y}}\Bigg) \numberthis
.\end{align*}
The distribution in Eq.~\eqref{eq:prob_dist} is the main result of this work and corresponds to the most general form of the count-probability distribution for the setup in Fig.~\ref{fig:setup}.

\begin{figure}[t]
	\centering
	\includegraphics{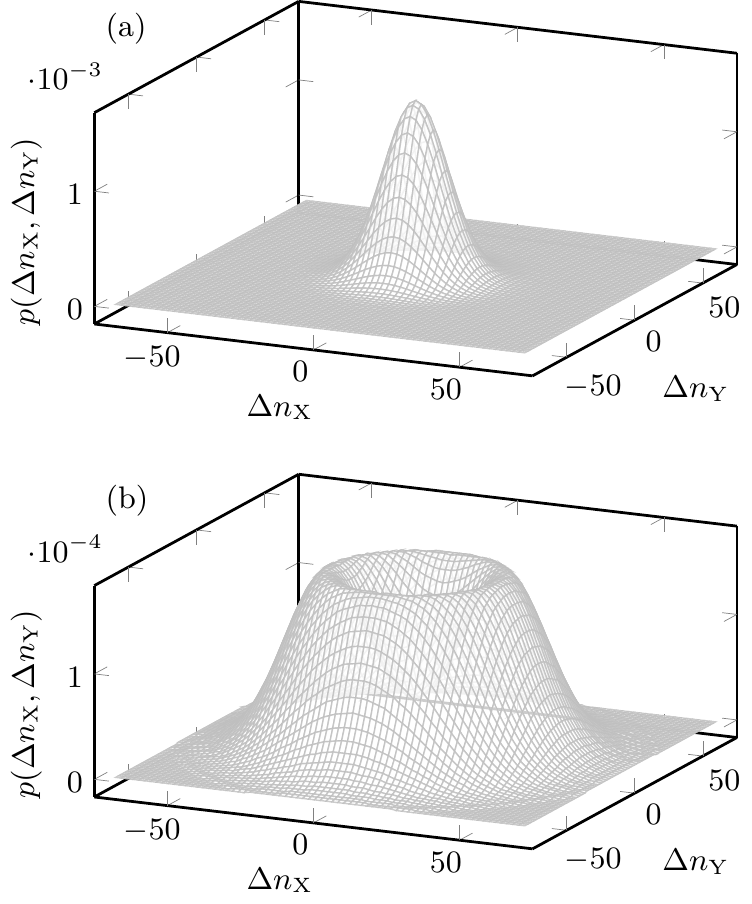}
	\caption{An example of the count-probability distribution $p(\Delta n_\text{X},\Delta n_\text{Y})$ corresponding to a symmetric $\overline{\text{XY}}$-measurement with equal pump strengths $\abs{\tilde{\alpha}_\text{X}} = \abs{\tilde{\alpha}_\text{Y}} = 1/\sqrt{2}$ and probes $\beta_\text{X} = \beta_\text{Y} = 10$ of a three-photon MIR Fock state $n_\Omega = 3$. In (a) the squeezing strength is assumed to be $\zeta = 0.1$ and in (b) it is assumed to be $\zeta = 1$. Although the photon-number differences are discrete, the distribution is broad enough to justify the use of a continuous surface plot.}
	\label{fig:example_fock}
\end{figure}

Let us highlight the special case with one $\hat{X}$ measurement and one $\hat{Y}$ measurement, or in other words $\abs{I_\text{X}} = \abs{I_\text{Y}} = 1$, and equal combined pump amplitudes ($A_X = A_Y$) for the $\hat{X}$ and $\hat{Y}$ measurements. We will call this measurement a symmetric  $\overline{\text{XY}}$-measurement from now on. For this configuration, the renormalization envelope $N(\{\Delta n_i\}) = (2\abs{\nu}^2\abs{\beta_{1}}\abs{\beta_{2}})^{-1}$ is constant. An example of the corresponding count-probability distribution, with a three-photon Fock state as an MIR input, can be seen in Fig.~\ref{fig:example_fock}.
\noindent
For weak squeezing, $\zeta = 0.1$, the distribution is dominated by the vacuum in the NIR modes, while for stronger squeezing, $\zeta=1$, the contribution of the MIR state becomes much clearer. This setup is similar to eight-port homodyne detection as considered by Freyberger \textit{et al.}~\cite{Freyberger_1993}. The approximation used in \cite{Freyberger_1993} is a special case of the Gaussian approximation for the Skellam distribution. While the two approximations coincide for the symmetric $\overline{\text{XY}}$-measurement, the former breaks down for a simultaneous measurement of more than two quadratures. The approximation of the Skellam distribution on the other hand is valid for arbitrarily many measurements. On top of that, it gives a physical intuition of the processes at the ellipsometry stage, as explained in section~\ref{s:count_prob}. In eight-port homodyne detection, the probability distribution is related to the Husimi function $\rho(z;s=-1)$ \cite{Husimi_1940}, as long as perfect detectors are considered. This corresponds to the limit of infinite squeezing: $\lim_{\abs{\zeta}\to\infty} \tilde{s} = -1$. If the detectors have a quantum efficiency $\eta < 1$ or the first beam splitter in the eight-port homodyne setup is replaced by a down-conversion process, the parameter $\tilde{s}$ of the sampled quasiprobability distribution can also be smaller than $-1$ \cite{Leonhardt_1993,Zucchetti_1996,Tiedau_2018}. In contrast to homodyne detection, electro-optic sampling is an indirect measurement. This allows not only to tune the strength of the measurement (so that back action on the sampled MIR state can be kept at a desired level \cite{Guedes_2022}), but also to investigate the post-measurement state in the MIR, since the indirect sampling is by nature non-destructive. The two measurement schemes are compared further in section~\ref{s:comparison}.

\subsection{Ensemble limit}\label{s:ensemle}
The probability distribution~\eqref{eq:prob_dist} can be used to calculate the expected photon-number differences $\ev{\Delta n_i}$, as well as the corresponding variances $\var{\Delta n_i}$. Let us consider the symmetric $\overline{\text{XY}}$-measurement from the previous section. The expectation value for, e.g., the first photon-number difference is then 
\begin{equation}
	\ev{\Delta n_\text{X}} = \sum_{\Delta n_\text{X}, \Delta n_\text{Y} = -\infty}^\infty \Delta n_\text{X} p(\Delta n_\text{X}, \Delta n_\text{Y})
.\end{equation}
For a strong probe, such that $\abs{\beta_{i}}\abs{\nu}$ is large, the sum can be approximated by an integral: $\ev{\Delta n_\text{X}} \approx \sqrt{2}\abs{\nu}\abs{\beta_\text{X}} \int \Re(z) \rho(z;s) \dd z$. This is a phase-space average and can thus be related to the (ensemble) expectation values of the MIR $\hat{X}$ quadrature, so that for $\hat{X}_\Omega = \frac{1}{2}(\hat{a}_\Omega + \hat{a}_\Omega^\dagger)$ and $\hat{Y}_\Omega = \frac{\iu}{2}(\hat{a}_\Omega^\dagger - \hat{a}_\Omega)$ one has
\begin{equation}\label{eq:expecation_val}
	\ev{\Delta n_\text{Q}} \approx \sqrt{2}\abs{\nu}\abs{\beta_\text{Q}}\ev{\hat{Q}_\Omega}
,\end{equation}
with $Q = X,Y$. Similarly, the (root mean square) variances of the photon-number differences can be related to the variances of the quadratures, $\var{\hat{X}} = \ev*{\hat{X}_\Omega^2}-\ev*{\hat{X}_\Omega}^2$ and $\var{\hat{Y}} = \ev*{\hat{Y}_\Omega^2}-\ev*{\hat{Y}_\Omega}^2 $, according to
\begin{equation}\label{eq:variance}
	\var{\Delta n_\text{Q}} \approx 2\abs{\nu}^2\abs{\beta_\text{Q}}^2\left(\var{\hat{Q}} - \frac{\tilde{s}}{4}\right)
.\end{equation}
While the probe amplitudes $\beta_\text{Q}$ rescale the quasiprobability distribution, the Weierstrass transform smoothes it resulting in additional noise. The variance in Eq.~\eqref{eq:variance} consists of two contributions scaled by the same factor $\sqrt{2}\abs{\nu}\abs{\beta_\text{Q}}$ as the quadrature expectation value in Eq.~\eqref{eq:expecation_val}: the first contribution comes from the unperturbed variance of the MIR-state quadratures $\var{\hat{Q}}$, so that the product $\var{\hat{X}}\var{\hat{Y}} \geq 1/16$ is limited by Robertson's uncertainty relation~\cite{Robertson_1929}; the second contribution is proportional to the parameter $\tilde{s}$ and is independent of the sampled state. Thus the first contribution can be attributed to the preparation uncertainty while the second term relates to the measurement uncertainty. As shown by Werner~\cite{Werner_2016,Werner_2019}, these uncertainty relations coincide. In this case, they are both bounded by $1/16$; i.e., $\tilde{s}^2/16 \geq 1/16$, $\var{\hat{X}}\var{\hat{Y}} \geq 1/16$. This bound is optimal for the simultaneous measurement of a single observable, which is a function of generalized position and momentum \cite{Werner_2019}. The relation $(\var{\hat{X}} - \tilde{s}/4)(\var{\hat{Y}} - \tilde{s}/4) \geq 1/4$ also agrees with the result from Arthurs and Goodman \cite{Arthurs_1988} about balanced homodyne detection.

\subsection{Marginal distributions}
The marginal distributions of the Wigner function $\rho(z;s=0)$ give the correct quantum mechanical probabilities for the generalized coordinate and conjugate momentum \cite{Wigner_1932}. Thus, we express the $(s_\text{X},s_\text{Y})$-parametrized quasiprobability distribution for $s_\text{X},s_\text{Y} < 0$ as
\begin{align*}\label{eq:qpd_wigner}
	&\rho(z;s_\text{X},s_\text{Y}) = \frac{1}{\pi}\frac{2}{\sqrt{s_\text{X}s_\text{Y}}} \int \rho(y;0) \\
	&\times \e^{2\Re^2(y-z)/s_\text{X}+2\Im^2(y-z)/s_\text{Y}} \dd y \numberthis
.\end{align*}
If only $\hat{X}$ ($\hat{Y}$) measurements are considered, the opposite-quadrature parameter, $\tilde{s}_\text{Y}$ ($\tilde{s}_\text{X}$), goes to $-\infty$ as the pump amplitude for the $\hat{Y}$ ($\hat{X}$) measurement vanishes: $\abs{\tilde{\alpha}_{i_\text{Y}}}\to 0$ ($\abs{\tilde{\alpha}_{i_\text{X}}} \to 0$). Thus, the probability distribution of sole $\hat{X}$ measurements is related to the marginal distribution of the Wigner function $\prescript{}{\Omega}{\bra{x}}\hat{\rho}_\Omega\ket{x}_\Omega$: 
\begin{align*}\label{eq:prb_dist_marginal}
	&p(\{\Delta n_i\}) \approx N_\text{X}(\{\Delta n_i\}) \int \prescript{}{\Omega}{\bra{x}}\hat{\rho}_\Omega\ket{x}_\Omega \\
	&\e^{\frac{2}{\tilde{s}_\text{X}}\left\{x-\Re\left[z(\{\Delta n_i\})\right]\right\}^2} \dd x \numberthis
,\end{align*}
with the renormalization envelope
\begin{align*}
	&N_\text{X}(\{\Delta n_i\}) = \frac{\sqrt{1-\tilde{s}_\text{X}}/\sqrt{-\tilde{s}_\text{X}}}{\mu^2\sqrt{1-\mu^4/\abs{\nu}^4}} \prod_{i \in I}\frac{\e^{-\Delta n_i^2/(2\abs{\beta_i}^2)}}{\sqrt{2\pi\abs{\beta_i}^2}} \\
	&\times \exp\left\{\left[1 + 2\frac{\nu^2+\mu^2}{1-\tilde{s}_\text{X}}\right]\frac{2\Re^2\left[z(\{\Delta n_i\})\right]}{1-\tilde{s}_\text{X}}\right\} \numberthis
.\end{align*}
Analogously, the probability distribution for only $\hat{Y}$ measurements is related to the quantum mechanical distribution of the $\hat{Y}$-quadrature, $\prescript{}{\Omega}{\bra{y}}\hat{\rho}_\Omega\ket{y}_\Omega$, with the roles of $X$ and $Y$ interchanged and $\Re\left[z(\{\Delta n_i\})\right]$ replaced by $\Im\left[z(\{\Delta n_i\})\right]$.

The count-probability distribution is not directly given by the quantum-mechanical distribution, but by its Weierstrass transform, resulting in a distribution smoothing that manifests itself as additional noise. 

\section{Post-measurement quasiprobability distribution}\label{s:post}
\begin{figure}[t]
	\centering
	\includegraphics{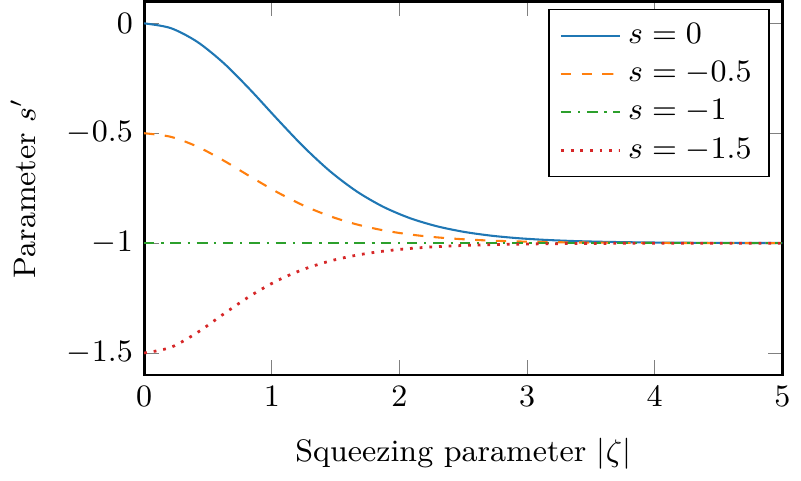}
	\caption{The parameter $s^\prime = s_\text{X}^\prime = s_\text{Y}^\prime$ of Eq.~\eqref{eq:post_msnt_qpd} for a setup with $A_X = A_Y$ as a function of the squeezing strength $\abs{\zeta}$ for different parameters of the post-measurement quasiprobability distribution $s = s_\text{X} = s_\text{Y}$. Among these are the Wigner function with $s = 0$ and the Husimi function with $s = -1$.}
	\label{fig:order_param_post}
\end{figure}

The state in the MIR mode after the measurement is
\begin{equation}\label{eq:post_msnt_state}
	\hat{\rho}_\Omega^\prime = \frac{1}{p(\{\Delta n_i\})} \sum_{\{n_i\}} \hat{M}_{\{n_i,\Delta n_i\}}\hat{\rho}_\Omega\hat{M}_{\{n_i,\Delta n_i\}}^\dagger
,\end{equation}
with the measurement operator
\begin{equation}\label{eq:povm_elements}
	\hat{M}_{\{n_i,\Delta n_i\}} = \bigotimes_{i \in I}\prescript{}{i,s}{\bra{n_i + \Delta n_i}}\prescript{}{i,z}{\bra{n_i}}\hat{U}\ket{0}_\omega
.\end{equation}
The summation over the photon-numbers $n_i$ of the $z$-polarized modes in Eq.~\eqref{eq:post_msnt_state} is necessary because EOS is a (partially) non-selective measurement. It is only selective with respect to the photon-number differences between the $s$- and $z$-polarized output channels, but not with respect to the actual photon numbers in each polarization. Using again the approximation of the Skellam distribution as already done for the count-probability distribution (see Appendix~\ref{a:post_meas} for details), the $(s_\text{X},s_\text{Y})$-parametrized post-measurement quasiprobability distribution
\begin{align*}\label{eq:post_msnt_qpd}
	&\rho^\prime(z;s_\text{X},s_\text{Y}) \\
	&= \frac{1}{\pi^2} \int \e^{z\conj{\gamma} - \conj{z}\gamma} \e^{\frac{1+s_\text{Y}}{2}\Re^2(\gamma) + \frac{1+s_\text{X}}{2}\Im^2(\gamma)} \chi^\prime(\gamma;-1) \dd \gamma \\
	&\approx N^\prime(z;s_\text{X},s_\text{Y}) \rho(z^\prime(z);s_\text{X}^\prime,s_\text{Y}^\prime) \numberthis
,\end{align*}
with $s_\text{Q} < 2\mu^2/(1+A_Q)-1$ and $Q=X,Y$, is given by the $(s_\text{X}^\prime,s_\text{Y}^\prime)$-parametrized quasiprobability distribution of the initial state $\hat{\rho}_\Omega$ and a renormalization envelope
\begin{align*}\label{eq:renorm_prime}
	&N^\prime(z;s_\text{X}^\prime,s_\text{Y}^\prime) = \frac{\sqrt{1-s_\text{X}^\prime}\sqrt{1-s_\text{Y}^\prime}}{2}\left(\prod_{i \in I} \frac{\e^{-\Delta n_i^2/(2\abs{\beta_{i}}^2)}}{\sqrt{2\pi \abs{\beta_{i}}^2}}\right)  \\
	&\times \left[\mu^2\sqrt{\frac{\mu^2}{1+A_X} - \frac{1+s_\text{X}}{2}}\sqrt{\frac{\mu^2}{1+A_Y} - \frac{1+s_\text{Y}}{2}}p(\{\Delta n_i\})\right]^{-1} \\
	&\times \exp\Bigg\{\frac{2\Re^2\left[z^\prime(z)\right]}{1-s_\text{X}^\prime} + \frac{2\Im^2\left[z^\prime(z)\right]}{1-s_\text{Y}^\prime} + \abs{\tilde{z}}^2 \\
	&-\Re^2(z-\tilde{z})\left(\frac{\mu^2}{1+A_X}-\frac{1+s_\text{X}}{2}\right)^{-1} \\
	&- \Im^2(z-\tilde{z})\left(\frac{\mu^2}{1+A_Y}-\frac{1+s_\text{Y}}{2}\right)^{-1}\Bigg\} \numberthis
.\end{align*}
Note that this envelope is different from the one in Eq.~\eqref{eq:renomrm_prob_dist}. The parameters of the pre-measurement distribution are 
\begin{align}
	s_\text{Q}^\prime = 1-2\mu^2\left[\abs{\nu}^2+\left(\frac{\mu^2}{1+A_{Q}} - \frac{1+s_\text{Q}}{2}\right)^{-1}\right]^{-1} \label{eq:s_prime}
,\end{align}
The parameters $s_\text{Q}$ can be freely chosen [as long as $s_\text{Q} < 2\mu^2/(1+A_Q)-1$ is fulfilled] and are not related to $\tilde{s}_\text{Q}$ from section~\ref{s:count_prob}. The restrictions on the $s_\text{Q}$ are necessary to ensure convergence of all integrals. Furthermore, the argument of the quasiprobability distribution $\rho$ of the initial state in Eq.~\eqref{eq:post_msnt_qpd} is rescaled and displaced according to
\begin{align*}\label{eq:z_prime}
	&z^\prime(z) = \frac{1-s_\text{X}^\prime}{2\mu}\Re\left(\tilde{z}+\frac{z-\tilde{z}}{\mu^2/(1+A_X)-(1+s_\text{X})/2}\right) \\
	&+\iu\frac{1-s_\text{Y}^\prime}{2\mu}\Im\left(\tilde{z}+\frac{z-\tilde{z}}{\mu^2/(1+A_Y)-(1+s_\text{Y})/2}\right) \numberthis
,\end{align*}
where the displacement is related to the measurement outcomes $\{\Delta n_i\}$ through $\tilde{z} = \frac{\abs{\nu}}{\mu}\tilde{y}$ with
\begin{equation}\label{eq:z_tilde}
	\tilde{y} = \sum_{i_\text{X} \in I_\text{X}}\frac{\abs{\tilde{\alpha}_{i_\text{X}}}}{\abs{\beta_{i_\text{X}}}}\Delta n_{i_\text{X}} + \iu\sum_{i_\text{Y} \in I_\text{Y}}\frac{\abs{\tilde{\alpha}_{i_\text{Y}}}}{\abs{\beta_{i_\text{Y}}}}\Delta n_{i_\text{Y}}
.\end{equation}
For a configuration with $A_X=A_Y$ the parameter $s^\prime = s_\text{X}^\prime = s_\text{Y}^\prime$ is shown in Fig.~\ref{fig:order_param_post} as a function of the squeezing strength for different values of the post-measurement quasiprobability distributions parameter $s = s_\text{X} = s_\text{Y}$. At the limit of very weak squeezing, $s^\prime$ tends to $s$-dependent plateaus due to the vanishing coupling between the MIR and the NIR modes. In the limit of strong squeezing, $s^\prime$ tends to $-1$ independently of the $s$ value of the post-measurement quasiprobability distribution; this means that the stronger the squeezing is, the more positive the quasiprobability distribution $\rho^\prime(z,s_\text{X}^\prime,s_\text{Y}^\prime)$ becomes, because it tends towards the Husimi function $\rho(z;s=-1)$.
\noindent
In fact, for infinite squeezing, the post-measurement quasiprobability distribution gives
\begin{align*}\label{eq:qpd_strong_limit}
	&\lim_{\abs{\zeta}\to\infty} \rho^\prime(z;0) \\
	&=\frac{2}{\pi}\exp\left[-2\e^{2r}\Re^2(z-\tilde{y})-2\e^{-2r}\Im^2(z-\tilde{y})\right] \numberthis
\end{align*}
and corresponds to a displaced squeezed state $\ket{\tilde{y},r}$ with squeezing parameter $r = \frac{1}{2}\ln\left(\sum_{i_\text{X}\in I_\text{X}} \abs{\tilde{\alpha}_{i_\text{X}}}^2/\sum_{i_\text{Y}\in I_\text{Y}} \abs{\tilde{\alpha}_{i_\text{Y}}}^2\right)$. This happens as a consequence of the monotonic dependence of the $\rho$ broadening on the squeezing parameter: $\lim_{\abs{\zeta}\to\infty} z^\prime(z) = 0$. For $A_X = A_Y$ in Eq.~\eqref{eq:qpd_strong_limit}, $r=0$, resulting in a coherent state. If only $\hat{X}$ ($\hat{Y}$) measurements are performed, $r\to\infty$($-\infty$) and Eq.~\eqref{eq:qpd_strong_limit} describes a $\hat{X}$-($\hat{Y}$-)quadrature eigenstate; in either case, the state defined by Eq.~\eqref{eq:qpd_strong_limit} does not explicitly depend on the initial quasiprobability distribution, but it does depend on the measurement outcomes $\{\Delta n_i\}$, which are conditioned by the choice of an initial state. This is in accordance with the results obtained by Arthurs and Kelly for an ideal measurement \cite{Arthurs_Kelly_1965,She_1966} as well as for a weak Arthurs-Kelly measurement \cite{Ochoa_2018}.

\begin{figure}[t]
	\centering
	\includegraphics{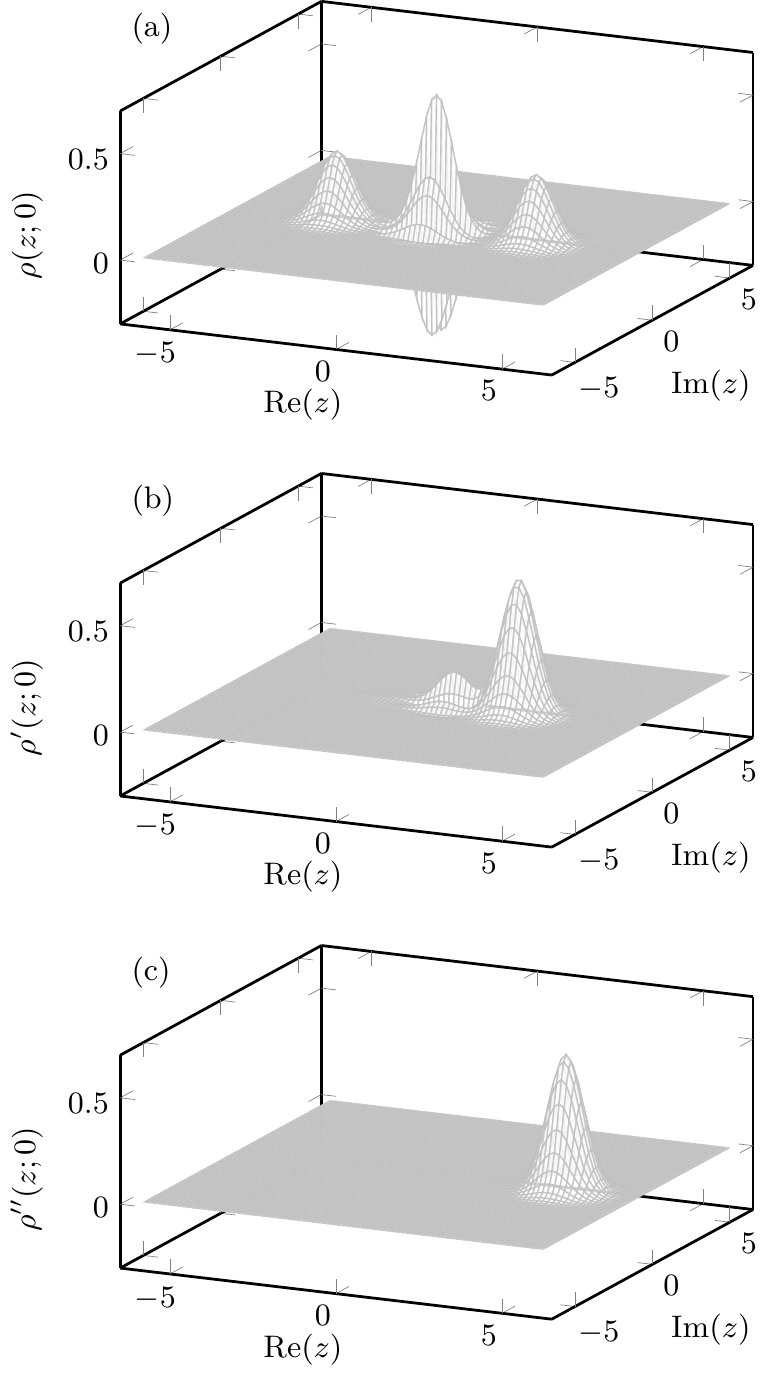}
	\caption{(a) Wigner function $\rho(z;0)$ of the cat state $\propto \ket{\alpha_\Omega} + \ket{-\alpha_\Omega}$ with $\alpha_\Omega = 3$. (b) Wigner function $\rho^\prime(z;0)$ of the cat state after a symmetric $\overline{\text{XY}}$-measurement is performed, $\zeta = 1$, $\beta_\text{X} = \beta_\text{Y} = 10$ and measurement outcomes $\Delta n_\text{X} = 10$, $\Delta n_\text{Y} = 0$. (c) Wigner function after yet another measurement with the outcomes $\Delta n_\text{X}^\prime = 40$, $\Delta n_\text{Y}^\prime = 0$. The outcomes of all measurements are not exceedingly small with $p_1(10,0) = 0.000012$ and $p_2(40,0) = 0.00064$.}
	\label{fig:wigner_post}
\end{figure}

The same reasoning applies for the case of infinitely many consecutive measurements. A consecutive measurement is defined as an EOS of the post-measurement MIR state using a copy of the setup in Fig.~\ref{fig:setup}. With every consecutive measurement, the argument of the quasiprobability distribution is rescaled by a factor $\mu^{-1}(1-s^\prime)/(1-s)$, which is smaller than 1 for $s < 1$ and  $\abs{\zeta}>0$. Thus, in the limit of infinitely many consecutive measurements, the argument of the quasiprobability distribution $z^\prime(z)$ tends to zero and the resulting quasiprobability distribution only depends on the renormalization envelopes. For a series of symmetric $\overline{\text{XY}}$-measurements, the final state seems to always tend to a coherent state if the measurement outcomes are not exceedingly small, as can be seen in Fig.~\ref{fig:wigner_post} for a cat state. This is also known to happen for a harmonic oscillator weakly coupled to a thermal bath~\cite{Zurek_1993}.

\section{Comparison of different simultaneous measurements}\label{s:comparison}
As a benchmark, the following problem is considered: Let us assume that the state of the MIR field is already categorized (e.g., squeezed, displaced, thermal, etc.) and can be parameterized with the parameters $\{\lambda_j\}$. If, for example, it is known that the MIR mode is in a coherent state $\ket{\alpha_\Omega}_\Omega$ with an unknown $\alpha_\Omega$, the task would be to determine the parameter $\lambda_1 = \alpha_\Omega$. The probability distribution of these undetermined parameters, $p(\{\lambda_j\} | \{\Delta n_i\})$, can be directly obtained from the count probability $p(\{\Delta n_i\}) = p(\{\Delta n_i\} | \hat{\rho}_\Omega) = p(\{\Delta n_i\} | \{\lambda_j\})$, since the latter is defined as the distribution of conditional probabilities to measure the photon-number difference $\Delta n_i$ given that the MIR mode is in the state $\hat{\rho}_\Omega$. Hence, the parameter probability distribution can be calculated using Bayes' theorem \cite{Buzek_1998},\cite[p.~42]{Bernardo_Smith}
\begin{align*}\label{eq:bayes}
	&p(\{\lambda_j\} | \{\Delta n_i\}) \\
	&= \frac{p(\{\Delta n_i\} | \{\lambda_j\}) p(\{\lambda_j\})}{\int p(\{\Delta n_i\} | \{\lambda_j\}) p(\{\lambda_j\})\prod_j\dd\mu_j(\lambda_j)} \numberthis
,\end{align*}
where $\mu_j$ is an appropriate measure for the parameter $\lambda_j$. Since there is no a priori information about the parameters $\{\lambda_j\}$, the initial parameter distribution $p(\{\lambda_j\})$ is assumed to be uniform, in accordance with the indifference principle. With this assumption, the parameter distribution can be expressed as
\begin{equation}
	p(\{\lambda_j\} | \{\Delta n_i\}) = \frac{\rho(z,\tilde{s}_\text{X},\tilde{s}_\text{Y})}{\int \rho(z,\tilde{s}_\text{X},\tilde{s}_\text{Y})\prod_j\dd\mu_j(\lambda_j)}
.\end{equation}
A simple, but for our purposes sufficient, reconstructed state reads~\cite{Buzek_1998}
\begin{equation}\label{eq:recunstructed}
	\hat{\rho}_\text{rec} = \int p(\{\lambda_j\} | \{\Delta n_i\})\hat{\rho}_{\{\lambda_j\}}\prod_j\dd\mu_j(\lambda_j)
,\end{equation}
where $\hat{\rho}_{\{\lambda_j\}}$ is the state parameterized by $\{\lambda_j\}$. For additional measurements, the Bayesian update can be repeated by replacing $p(\{\lambda_j\})$ with $p(\{\lambda_j\} | \{\Delta n_i\})$ as a better estimate of the actual distribution of the parameters \cite[p.48]{Bernardo_Smith}. We will denote the count-probability distribution of the first EOS measurement as $p_1(\{\Delta n_i\} | \{\lambda_j\})$. The count-probability distribution of the second EOS measurement will be denoted as $p_2(\{\Delta n_{i^\prime}^\prime\}|\{\lambda_j\})$, where $\Delta n_{i^\prime}^\prime$ are the measurement outcomes of the second EOS with the NIR frequencies $i^\prime \in I^\prime$. Hence, applying Bayes law again will update the parameter distribution to
\begin{align*}\label{eq:bayes_2}
	&p(\{\lambda_j\} | \{\Delta n_i\}, \{\Delta n_{i^\prime}^\prime\}) \\
	&= \frac{p_2(\{\Delta n_{i^\prime}^\prime\}|\{\lambda_j\})p_1(\{\lambda_j\}|\{\Delta n_i\})}{\int p_2(\{\Delta n_{i^\prime}^\prime\}|\{\lambda_j\})p_1(\{\lambda_j\}|\{\Delta n_i\})\prod_j \dd \mu_j(\lambda_j)} \numberthis
.\end{align*}
The reconstructed state is thus
\begin{equation}\label{eq:recunstructed_2}
	\hat{\rho}_\text{rec} = \int p(\{\lambda_j\} | \{\Delta n_i\}, \{\Delta n_{i^\prime}^\prime\})\hat{\rho}_{\{\lambda_j\}}\prod_j\dd\mu_j(\lambda_j)
.\end{equation}
The fidelity $F(\hat{\rho}_\Omega,\hat{\rho}_\text{rec})$ can be used to quantify how close the reconstructed state is to the initial state \cite[p.~409-423]{Nielsen_2009}. However, this fidelity depends on the measurement outcomes $\{\Delta n_i\}$. To get a quantity independent of the measurement outcomes, the average fidelity $\ev{F(\hat{\rho}_\Omega,\hat{\rho}_\text{rec})} = \sum_{\{\Delta n_i\}} F(\hat{\rho}_\Omega,\hat{\rho}_\text{rec}) p(\{\Delta n_i\})$ can be used.

Similarly, the effect of the measurement on the MIR mode is quantified by the fidelity $F(\hat{\rho}_\Omega,\hat{\rho}_\Omega^\prime)$ between the initial and the post-measurement state. If the initial state is pure, the fidelity
\begin{equation}\label{eq:fidelity}
	F(\hat{\rho}_\Omega,\hat{\rho}_\Omega^\prime) = \pi \int \rho(z;0) \rho^\prime(z;0) \dd z
\end{equation}
can be expressed using the Wigner function of the initial state $\rho(z;0)$ and post-measurement state $\rho^\prime(z;0)$. Once again, to get a measurement-outcome independent quantity, the average fidelity is considered. 
\begin{figure}[t]
	\centering
	\includegraphics{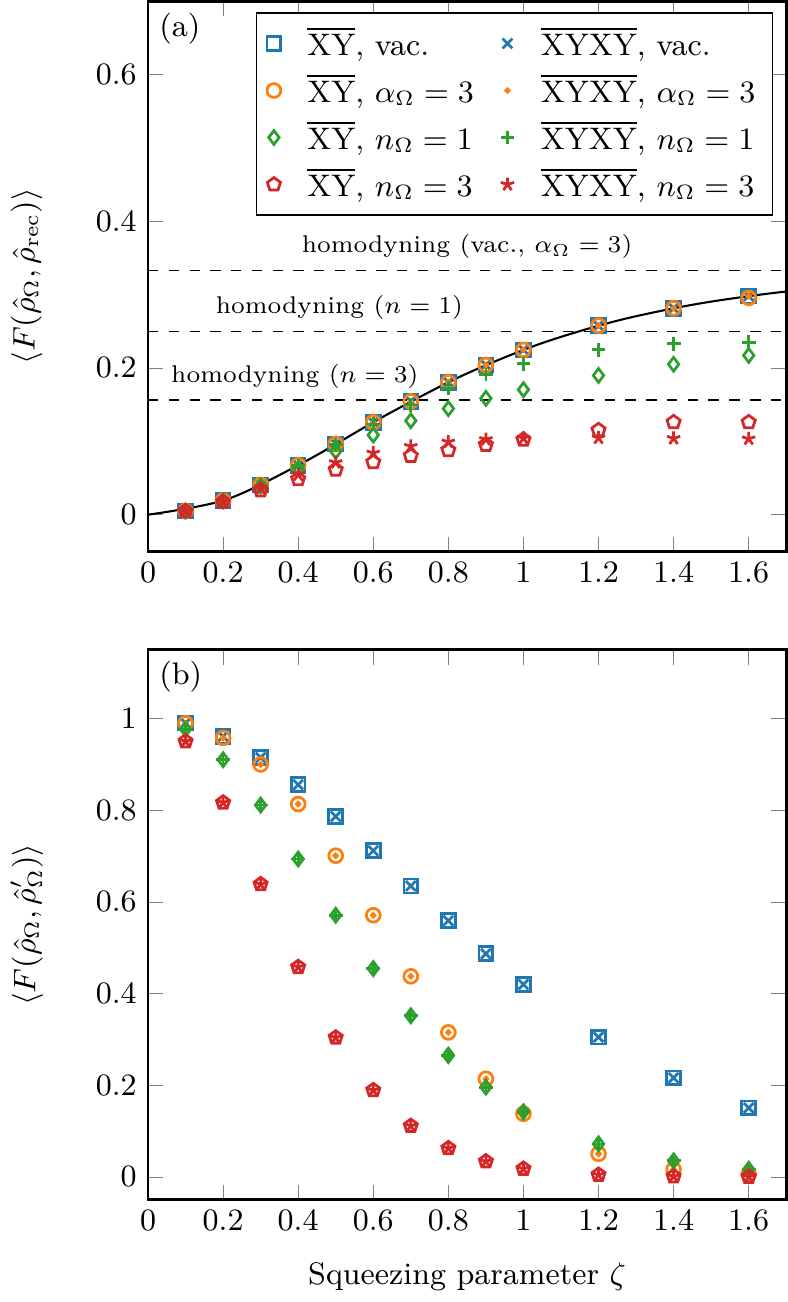}
	\caption{(a) Numerical evaluation of the average fidelity between the initial MIR state and the reconstructed state in Eq.~\eqref{eq:recunstructed} for various squeezing parameters. The initial state was assumed to be the vacuum, a coherent state with $\alpha_\Omega=3$ and two Fock states with $n_\Omega = 1$ as well as $n_\Omega = 3$ photons. Also two different measurement setups where considered. A symmetric $\overline{\text{XY}}$-measurement, as well as a symmetric $\overline{\text{XYXY}}$-measurement. The dashed horizontal lines indicate the fidelity if eight-port homodyne detection would be used in place of electro-optic sampling. The solide line is an approximate solution. (b) Average fidelity between the initial and the post-measurement state of the MIR field. The initial states are the same as in (a).}
	\label{fig:comparison_sim}
\end{figure}
\begin{figure}[t]
	\centering
	\includegraphics{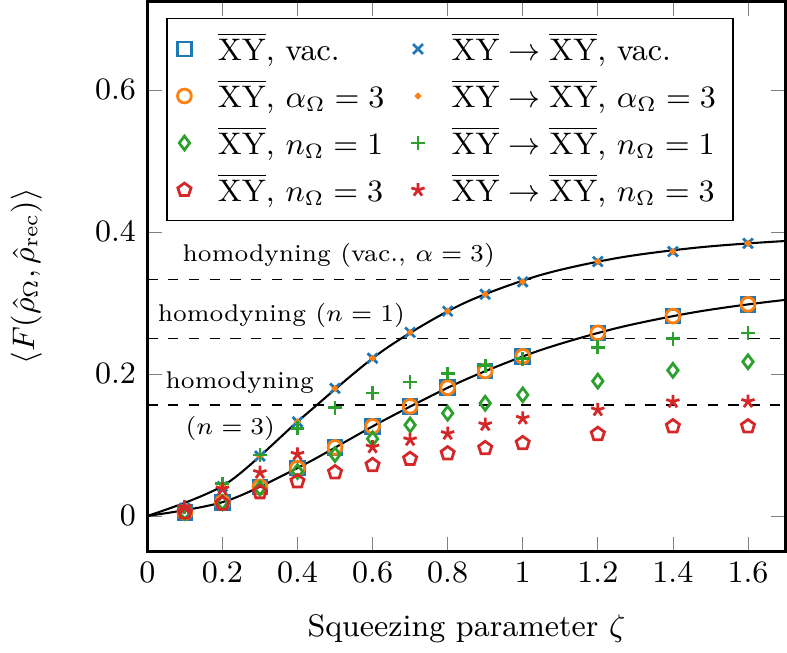}
	\caption{Numerical evaluation of the average fidelity between the initial and the reconstructed state for various squeezing parameters. The initial state was assumed to be the vacuum, a coherent state with $\alpha_\Omega=3$ and two Fock states with $n_\Omega = 1$ as well as $n_\Omega = 3$ photons. Also two different measurement setups where considered. A symmetric $\overline{\text{XY}}$-measurement and a consecutive measurement $\overline{\text{XY}} \rightarrow \overline{\text{XY}}$ (a second $\overline{\text{XY}}$ electro-optic measurement on the post-measurement state). The dashed horizontal lines indicate the fidelity if eight-port homodyne detection would be used in place of electro-optic sampling. The solid lines represent an approximate solution.}
	\label{fig:comparison_cons}
\end{figure}
\noindent
Numerical results for the average fidelities between the initial and the reconstructed states of a symmetric $\overline{\text{XY}}$-measurement and a symmetric $\overline{\text{XYXY}}$-measurement can be seen in Fig.~\ref{fig:comparison_sim} (a), while the corresponding average fidelities between the initial and post-measurement states is shown in Fig.~\ref{fig:comparison_sim} (b). A symmetric $\overline{\text{XYXY}}$-measurement is a simultaneous electro-optic measurement of two $\hat{X}$ quadratures and two $\hat{Y}$ quadratures with equal combined pump strength for both observables, $A_X = A_Y$. The initial states are considered to be: the vacuum, a coherent state with $\alpha_\Omega = 3$, and two Fock states with photon-numbers $n_\Omega = 1$ and $n_\Omega = 3$. For the fidelity between an initially coherent state and the reconstructed state, an approximate analytical solution has also been derived (cf. appendix~\ref{ap:fidelity}). The fidelities between initial and reconstructed states obtained from electro-optic sampling converge in the limit of infinitely strong squeezing to the fidelities corresponding to eight-port homodyne detection. This agrees with the limit $\lim_{\abs{\zeta} \to \infty} \tilde{s} = -1$ discussed in section~\ref{s:count_prob}. The same holds for the symmetric $\overline{\text{XYXY}}$-measurement. In fact, the fidelities of the two measurement schemes coincide for the coherent states. For the Fock states there is a small deviation, not enough to overcome the limit set by eight-port homodyne detection; this happens because the renormalization envelope~\eqref{eq:renomrm_prob_dist} restricts the pair $(\Delta n_{\text{X}_1}, \Delta n_{\text{Y}_1})$ of one $\overline{\text{XY}}$-measurement to an area close the pair $(\Delta n_{\text{X}_2}, \Delta n_{\text{Y}_2})$ of the other $\overline{\text{XY}}$-measurements. This agrees with the results obtained by Braunstein \textit{et al.} \cite{Braunstein_1991}. The small deviation could originate from the fact that the quasiprobability distribution is only sampled discretely and thus different $z(\{\Delta n_i\})$ could lead to different fidelities. As a general trend, the fidelities between the initial and the post-measurement states are close to one for small squeezing and drop to zero for stronger squeezing, as expected because stronger measurements tend to disturb the measured state to a higher extent. Furthermore, the fidelities for Fock states drop faster than the ones for coherent states.

There is another way to obtain two $\hat{X}$- and two $\hat{Y}$-quadrature measurements. First a symmetric $\overline{\text{XY}}$-measurement is performed, then the same measurement is done on the post-measurement state of the first EOS (i.e., a consecutive measurement). We will denote this by $\overline{\text{XY}} \rightarrow \overline{\text{XY}}$. To obtain the parameter distribution, and therewith the fidelity corresponding to the reconstructed state, the Bayesian update has to be initially performed with the probability distribution of the first measurement $p_1(\{\Delta n_i\} | \{\lambda_j\})$ and then again with the probability distribution of the second measurement $p_2(\{\Delta n_{i^\prime}^\prime\} | \{\lambda_j\})$ resulting in Eq.~\eqref{eq:bayes_2}. A numerical evaluation of these fidelities can be seen in Fig.~\ref{fig:comparison_cons}. One can see that the fidelities between the initial and the reconstructed states exceed the limits set by eight-port homodyne detection. For a $\overline{\text{XY}}$-measurement of a coherent state, the asymptote of the approximate solution is $1/3$, while for the consecutive case it improves to $2/5$. This asymptotic behavior might appear to be inconsistent with the post-measurement state in the strong squeezing limit, Eq.~\eqref{eq:qpd_strong_limit}, since the quasiprobability distribution after the measurement seems to be independent of the initial state; however, while the post-measurement quasiprobability distribution does not depend on the quasiprobability distribution of the initial state, it does depend on the outcomes $\{\Delta n_i\}$ of the first measurement, which are conditioned to the initial state and its phase-space representations. Hence, the $\zeta\to\infty$ post-measurement state is not completely independent of the initial state and allows the consecutive-measurement fidelity between the initial and the reconstructed states to exceed the limit of a $\overline{\text{XY}}$-measurement.

\section{Conclusions}
In this work, we propose a multi-channel version of electro-optic sampling involving monochromatic modes and describe such measurements within the framework of the POVM formalism. In this scheme, the MIR mode interacts with multiple NIR pump modes, allowing for arbitrary combinations of $\hat{X}$- and/or $\hat{Y}$-quadrature measurements. The count-probability distribution was shown to be determined by a quasiprobability distribution together with a renormalization envelope. The parameters $\tilde{s}_\text{X},\tilde{s}_\text{Y}$ of this quasiprobability distribution only depend on the parameters associated with the nonlinear interaction. These parameters $\tilde{s}_\text{X},\tilde{s}_\text{Y}$, and thus the nonlinear interaction, determine an extra noise term on top of the quadrature variances (i.e., the shot-noise), while the probe only rescales the distribution. If only $\hat{X}$ ($\hat{Y}$) measurements are performed, the count-probability is related to the Weierstrass transform of the quantum-mechanical quadrature distribution $\ev{\hat{\rho}_\Omega}{x}$ ($\ev{\hat{\rho}_\Omega}{y}$).

For the post-measurement quasiprobability distributions, the effect of the measurement is a change in the quasiprobability distributions parameters, a renormalization and shift of both the quasiprobability distribution and its argument. In the limit case of an infinitely strong nonlinear interaction, the quasiprobability distribution represents a displaced squeezed state; if the sum of all NIR pump amplitudes of the $\hat{X}$ measurements equals that of all $\hat{Y}$ measurements, it represents a coherent state, while the post-measurement state is squeezed into a quadrature eigenstate as either the $\hat{X}$ or $\hat{Y}$ quadrature measurements prevail over the other. 

Finally, we compared several variants of multi-channel electro-optic sampling with the well established eight-port homodyne detection.
No significant difference was found between the $\overline{\text{XY}}$-measurement and the (simultaneous) $\overline{\text{XYXY}}$-measurement. This can be explained by correlations between the first $\hat{X}$, $\hat{Y}$ pair and the second one. In the limit of infinite squeezing, the fidelity between the reconstructed state and the initial MIR state asymptotically tends to the value set by eight-port homodyne detection. Nevertheless, the multi-channel EOS can exceed this limit and therefore eight-port homodyne measurements by using two consecutive measurements of both quadratures, as we showed for a diverse selection of quantum states (see Fig.~\ref{fig:comparison_cons}).

The fidelity between the initial and reconstructed states could potentially be further optimized by varying all the parameters allowed by the present description. For example, a consecutive measurement with different squeezing strengths for each measurement could be considered, to minimize the measurement back-action. Additionally, the fidelity could be improved by simultaneously measuring an additional intermediate quadrature different from the $\hat{X}$ or the $\hat{Y}$ quadratures.

\begin{acknowledgments}
	We acknowledge funding by the Deutsche Forschungsgemeinschaft (DFG) - Project No. 425217212 - SFB 1432.
	T.L.M.G. gratefully acknowledges the funding by the Baden-Württemberg Stiftung via the Elite Programme for Postdocs. 
	
\end{acknowledgments}

\appendix
\section{Count-probability distribution}\label{a:count_prob}
In this seciton, we will use an alternative representation for the action of the wave plate on the NIR mode $i$ using the matrix notation:
\begin{equation}\label{eq:phase_plate_matrix}
	\hat{U}_{i,\text{WP}} \begin{pmatrix} 
	\hat{a}_{i,s} \\
	\hat{a}_{i,z}
	\end{pmatrix} \hat{U}_{i,\text{WP}}^\dagger =
	\e^{-\iu\phi_i/2}
	\begin{pmatrix} 
		(W_i)_{11} & (W_i)_{12} \\
		(W_i)_{21} & (W_i)_{22} \\
	\end{pmatrix}
	\begin{pmatrix} 
	\hat{a}_{i,s} \\
	\hat{a}_{i,z}
	\end{pmatrix}
,\end{equation}
with the matrix elements
\begin{align}
	&(W_i)_{11} = \conj{(W_i)_{22}} = \cos(\phi_i/2) + \iu\sin(\phi_i/2)\cos(2\theta_i), \\
	&(W_i)_{12} = (W_i)_{21} = \iu\sin(\phi_i/2)\sin(2\theta_i)
.\end{align}
Starting from Eq.~\eqref{eq:def_prob_dist} and using the Glauber-Sudarshan distributions $\rho_{n_i + \Delta n_i}(\eta_{i,s};1)$ and $\rho_{n_i}(\eta_{i,z};1)$ for the Fock states $\ket{n_i + \Delta n_i}_{i,s}$ and $\ket{n_i}_{i,z}$,

\begin{align*}%\label{eq:nir_glauber}
	&\hat{P}_{\Delta n_i} = \iint \sum_{n_i=\tilde{n}_i}^\infty \rho_{n_i + \Delta n_i}(\eta_{i,s};1)\rho_{n_i}(\eta_{i,z};1) \\
	&\times \ket{\eta_{i,s}}_{i,s}\prescript{}{i,s}{\bra{\eta_{i,s}}} \otimes \ket{\eta_{i,z}}_{i,z}\prescript{}{i,z}{\bra{\eta_{i,z}}} \dd\eta_{i,s} \dd\eta_{i,z} \numberthis
,\end{align*}
as well as for the MIR state [Eq.~\eqref{eq:thz_glauber}] we can express the count-probability distribution as a convolution of the Glauber-Sudarshan distributions with some matrix elements of the time-evolution operator, 

\begin{align*}
	&p(\{\Delta n_i\}) = \prod_{i \in I}\sum_{n_i=\tilde{n}_i}^\infty \iiiint \rho(\gamma;1) \rho_{n_i+\Delta n_i}(\eta_{i,s};1) \\
	&\times \rho_{n_i}(\eta_{i,z};1) \abs{\prescript{}{\Omega}{\bra{\gamma^\prime}}\prescript{}{i,s}{\bra{\eta_{i,s}}}\prescript{}{i,z}{\bra{\eta_{i,z}}}\hat{U}\ket{0}_{i,z}\ket{0}_{i,s}\ket{\gamma}_\Omega}^2\\
	&\times \dd \gamma \dd \gamma^\prime \dd \eta_{i,s} \dd \eta_{i,z} \numberthis
.\end{align*}
We rewrite the total time-evolution operator defined by Eq.~\eqref{eq:time_evolution} as
\begin{align*}
	&\hat{U} = \hat{U}_{\text{WP}}\hat{D}_z(\vec{\beta})\hat{U}_\text{NL} \\
	&= \hat{U}_{\text{WP}}\hat{D}_z(\vec{\beta})\hat{U}_{\text{WP}}^\dagger\hat{U}_{\text{WP}}\hat{U}_\text{NL}\hat{U}_{\text{WP}}^\dagger\hat{U}_{\text{WP}} \numberthis
.\end{align*}
This can later be simplified using $\hat{U}_{\text{WP}}\ket{0}_\omega = \ket{0}_\omega$, with the vacuum NIR state $\ket{0}_\omega = \bigotimes_{i \in I} \ket{0}_{i,s}\ket{0}_{i,z}$, and
\begin{align*}
	&\hat{U}_{\text{WP}}\hat{D}_{i,z}(\beta_i)\hat{U}_{\text{WP}}^\dagger \\
	&= \hat{D}_{i,s}\left[\conj{(W_i^{-1})_{21}}\beta_i\right]\otimes\hat{D}_{i,z}\left[\conj{(W_i^{-1})_{22}}\beta_i\right] \numberthis
.\end{align*}
The squeezing operator can also be simplified. First, we decompose it using $\mu = \cosh(\abs{\zeta})$ and $\nu = \frac{\zeta}{\abs{\zeta}}\sinh(\abs{\zeta})$ \cite[p.~100]{Vogel_Welsch}:
\begin{align*}
	&\hat{U}_\text{NL} = \exp(\conj{\zeta}\hat{a}_{\Omega,s}\sum_{i \in I} \tilde{\alpha}_i \hat{a}_{i,s}-\hc) \\
	&= \exp(-\frac{\nu}{\mu}\hat{a}_{\Omega,s}^\dagger\sum_{i \in I} \conj{\tilde{\alpha}_i} \hat{a}_{i,s}^\dagger) \\
	&\times \left(\frac{1}{\mu}\right)^{{\hat{a}_{\Omega,s}}^\dagger\hat{a}_{\Omega,s} + \left(\Sigma_{i \in I} \tilde{\alpha}_i \hat{a}_{i,s}\right)^\dagger\Sigma_{i \in I} \tilde{\alpha}_i \hat{a}_{i,s} + 1} \\
	&\times \exp(\frac{\conj{\nu}}{\mu}\hat{a}_{\Omega,s}\sum_{i \in I} \tilde{\alpha}_i \hat{a}_{i,s}) \numberthis
.\end{align*}
Now, all terms with $\hat{a}_{i,s}\ket{0}_\omega = 0$ or $\hat{a}_{i,s}\ket{0}_\omega = 0$ vanish when applied to the vacuum:
\begin{equation}
	\hat{U}_\text{NL}\ket{0}_\omega = \exp(-\frac{\nu}{\mu}\hat{a}_{\Omega,s}^\dagger\sum_{i \in I} \conj{\tilde{\alpha}_i} \hat{a}_{i,s}^\dagger)\left(\frac{1}{\mu}\right)^{\hat{a}_{\Omega,s}^\dagger\hat{a}_{\Omega,s} + 1}\ket{0}_\omega
.\end{equation}
Final, the matrix elements of the total time-evolution operator~\eqref{eq:time_evolution} in the coherent-state basis, using $\ket{\eta}_\omega = \bigotimes_{i \in I} \ket{\eta_{i,s}}_{i,s}\ket{\eta_{i,z}}_{i,z}$, can be calculated:
\begin{align*}
	&\prescript{}{\Omega}{\bra{\gamma^\prime}}\prescript{}{\omega}{\bra{\eta}}\hat{U}\ket{0}_{\omega}\ket{\gamma}_\Omega = \frac{1}{\mu}\e^{-\abs{\gamma}^2/2 - \abs{\gamma^\prime}^2/2 + \gamma\conj{(\gamma^\prime)}/\mu} \Bigg\{\prod_i \\
	&\times \prescript{}{i,s}{\bra{\eta_{i,s}}}\hat{D}_{i,s}\left[\conj{(W_i^{-1})_{21}}\beta_i\right] \e^{-\frac{\nu}{\mu} \conj{\tilde{\alpha}_i}\conj{(\gamma^\prime)}\conj{(W_i^{-1})_{11}}\hat{a}_{i,s}} \ket{0}_{i,s} \\
	&\times \prescript{}{i,z}{\bra{\eta_{i,s}}}\hat{D}_{i,z}\left[\conj{(W_i^{-1})_{22}}\beta_i\right] \e^{-\frac{\nu}{\mu} \conj{\tilde{\alpha}_i}\conj{(\gamma^\prime)}\conj{(W_i^{-1})_{12}}\hat{a}_{i,z}} \ket{0}_{i,z} \Bigg\} \\
	&=\e^{\Sigma_i -\frac{\nu}{\mu}\conj{\tilde{\alpha}_i}\conj{(\gamma^\prime)}\conj{(W_i^{-1})_{11}}(\eta_{i,s}-\conj{(W_i^{-1})_{21}}\beta_i) - \abs{\eta_{i,s}-\conj{(W_i^{-1})_{21}}}^2/2} \\
	&\times\e^{\Sigma_i -\frac{\nu}{\mu}\conj{\tilde{\alpha}_i}\conj{(\gamma^\prime)}\conj{(W_i^{-1})_{12}}(\eta_{i,z}-\conj{(W_i^{-1})_{22}}\beta_i) - \abs{\eta_{i,z}-\conj{(W_i^{-1})_{22}}}^2/2} \\
	&\frac{1}{\mu}\e^{-\abs{\gamma}^2/2 - \abs{\gamma^\prime}^2/2 + \gamma\conj{(\gamma^\prime)}/\mu} \\
	&=\e^{-\frac{\nu}{\mu}\conj{\tilde{\alpha}_i}(W_i)_{11}\conj{(\gamma^\prime)}\conj{\eta_{i,s}}-\abs{\eta_{i,s}-(W_i)_{21}\beta_{i}}^2/2} \\
	&\times \e^{-\frac{\nu}{\mu}\conj{\tilde{\alpha}_i}(W_i)_{12}\conj{(\gamma^\prime)}\conj{\eta_{i,z}}-\abs{\eta_{i,z}-(W_i)_{22}\beta_{i}}^2/2} \\
	&\times \frac{1}{\mu}\exp(-\abs{\gamma}^2/2-\abs{\gamma^\prime}^2/2+\gamma\conj{(\gamma^\prime)}/\mu) \numberthis
.\end{align*}

In the last equality, we used $\conj{(W_i^{-1})_{11}}(W_i^{-1})_{21} + \conj{(W_i^{-1})_{12}}(W_i^{-1})_{22} = 0$. Since the matrix elements are Gaussian functions, the convolutions give the Weierstrass transforms of $\rho_{n_i + \Delta n_i}(\eta_{i,s};1)$ and $\rho_{\Delta n_i}(\eta_{i,z};1)$ [i.e., the Husimi function $\rho_{n_i + \Delta n_i}(\sqrt{\widetilde{m}_{i,1}};-1)$ and $\rho_{n_i}(\sqrt{\widetilde{m}_{i,2}};-1)$]. The count distribution can thus be written as:
\begin{align*}\label{eq:count_prob_with_skellam}
	&p(\{\Delta n_i\}) = \frac{\pi}{\mu^2} \exp(-\sum_{i \in I}\abs{\beta_{i}}^2) \int \rho(\gamma;1) \e^{-\abs{\gamma}^2} \\
	&\times \int \exp[-\abs{\gamma^\prime}^2+\frac{2}{\mu}\Re(\gamma^\prime\conj{\gamma}) + \sum_{i \in I} \widetilde{m}_{i,1} + \widetilde{m}_{i,2}] \\
	&\times \prod_{i \in I}\sum_{n_i=\tilde{n}_i}^\infty \rho_{n_i + \Delta n_i}(\sqrt{\widetilde{m}_{i,1}};-1)\rho_{n_i}(\sqrt{\widetilde{m}_{i,2}};-1) \dd\gamma^\prime \dd\gamma\numberthis
,\end{align*}
with
\begin{equation}\label{eq:phase_1}
	\widetilde{m}_{i,j} = \abs{\conj{(W_i)_{2j}}\beta_{i}-\frac{\nu}{\mu}\conj{\tilde{\alpha}_i}(W_i)_{1j}\conj{(\gamma^\prime)}}^2
.\end{equation}
The last line in Eq.~\eqref{eq:count_prob_with_skellam} can be expressed using the Skellam distribution \cite{Skellam_1946},
\begin{align*}\label{eq:skellam_dist}
	&p_\text{s}(\Delta n_i; \widetilde{m}_{i,1}, \widetilde{m}_{i,2}) \\
	&=\sum_{n_i=\tilde{n}_i}^\infty \rho_{n_i + \Delta n_i}(\sqrt{\widetilde{m}_{i,1}};-1)\rho_{n_i}(\sqrt{\widetilde{m}_{i,2}};-1) \\
	&= \e^{-(\widetilde{m}_{i,1}+\widetilde{m}_{i,2})}\left(\frac{\widetilde{m}_{i,1}}{\widetilde{m}_{i,2}}\right)^{\Delta n_i/2} I_{\Delta n_i}(2\sqrt{\widetilde{m}_{i,1} \widetilde{m}_{i,2}}) \numberthis
,\end{align*}
with the modified Bessel function $I$, which is the probability distribution of the difference $\Delta n_i$ between two Poissonian counting events with expectation values $\widetilde{m}_{i,1}$ and $\widetilde{m}_{i,2}$

To get a balanced signal (as discribed in the main text), the expectation value $\widetilde{m}_{i,1} - \widetilde{m}_{i,2}$ of the argument of the Skellam distribution has to be proportional to $\gamma^\prime$ [terms proportional to $\abs{\beta_i}^2$ would lead to a shift of $\Delta n_i$, as will become clear from Eq.~\eqref{eq:skellam_expect} and \eqref{eq:skellam_approx} later]. All the unwanted terms cancel if the matrix elements describing the effect of the wave plates are restricted to $(W_i)_{11}/(W_i)_{21} = -(W_i)_{12}/(W_i)_{22}$, leading to Eq.~\eqref{eq:eos_condition}. For $\frac{\pi}{2} \leq \phi_i \leq \frac{3}{2}\pi$, the fraction
\begin{align*}\label{eq:phase_2}
	&\frac{(W_i)_{11}}{(W_i)_{21}} = (-1)^{k_1+k_2} \left[\sqrt{-\cos(\phi_i)} - (-1)^{k_2} \iu\sqrt{2}\cos(\frac{\phi_i}{2})\right] \\
	&= \exp\Big\{(-1)^{k_2}\iu \arcsin[\sqrt{2}\cos(\phi_i/2)] + \iu\pi[k_1+k_2+1]\Big\} \numberthis
\end{align*}
is  a complex phase (because $\abs{(W_i)_{21}(W_i)_{22}} = \frac{1}{2}$, $\abs{(W_i)_{21}}/\abs{(W_i)_{22}} = 1$) and the count-probability distribution can be expressed as
\begin{align*}\label{eq:prob_dist_4}
	&p(\{\Delta n_i\}) = \frac{\pi}{\mu^2} \exp(-\sum_{i \in I}\abs{\beta_{i}}^2) \int \rho(\gamma;1) \e^{-\abs{\gamma}^2} \\
	&\times \int \exp[-\abs{\gamma^\prime}^2+\frac{2}{\mu}\Re(\gamma^\prime\conj{\gamma}) + \sum_{i \in I} m_{i,1} + m_{i,2}] \\
	&\times \prod_{i \in I} p_\text{s}(\Delta n_i; m_{i,1}, m_{i,2}) \dd\gamma^\prime \dd\gamma \numberthis
,\end{align*}
with
\begin{equation}\label{eq:phase_3}
	m_{i,j} = \frac{1}{2}\abs{\abs{\beta_{i}}+(-1)^{j-1}\e^{\iu\varphi_i}\frac{\abs{\nu}}{\mu}\abs{\tilde{\alpha}_i}\conj{(\gamma^\prime)}}^2 
.\end{equation}
The expectation values of the arguments of the Skellam distributions~$p_\text{s}(\Delta n_i; m_{i,1}, m_{i,2})$~\cite{Karlis_2006},
\begin{equation}\label{eq:skellam_expect}
	m_{i,1} - m_{i,2} = 2 \frac{\abs{\nu}}{\mu}\abs{\tilde{\alpha}_i}\abs{\beta_i}\Re\left[\e^{\iu\varphi_i}\conj{(\gamma^\prime)}\right],
\end{equation}
are proportional to $\gamma^\prime$. The variances
\begin{equation}\label{eq:skellam_var}
	m_{i,1} + m_{i,2} = \abs{\beta_{i}}^2 + \frac{\abs{\nu}^2}{\mu^2}\abs{\tilde{\alpha}_i}^2\abs{\gamma^\prime}^2 \approx \abs{\beta_{i}}^2
,\end{equation}
on the other hand, are approximately given by the probe amplitudes. The latter are assumed to be much stronger than the contribution from the nonlinear interaction. Since the variances~\eqref{eq:skellam_var} are large, the Skellam distributions can be approximated as normal distributions with the same expectation values and variances,
\begin{align*}\label{eq:skellam_approx}
	&p_\text{s}(\Delta n_i; m_{i,1}, m_{i,2}) \\
	&\approx \frac{1}{\sqrt{2\pi}\abs{\beta_{i}}} \e^{-\left\{\Delta n_i -2 \frac{\abs{\nu}}{\mu}\abs{\tilde{\alpha}_i}\abs{\beta_{i}}\Re\left[\e^{\iu\varphi_i}\conj{(\gamma^\prime)}\right]\right\}^2/(2\abs{\beta_{i}}^2)} \numberthis
.\end{align*}
This approximation is derived in the next section.

\section{Skellam distribution}\label{ap:skellam}
Let us assume that $\rho(\gamma;1) \approx 0$ for large $\gamma$, thus only small values for $\gamma$ contribute. Using that $\exp(-\abs{\gamma^\prime - \gamma/\mu}^2) \approx 0$ for large $\gamma^\prime$ and small $\gamma$, we can conclude that only small $\gamma^\prime$ are relevant and therefore $m_{i,1},m_{i,2}$ are large if $\beta_{i}$ is large, which is the case for strong probe amplitudes. This allows to approximate the Husimi function by a Gaussian distributions:
\begin{align*}
	&\pi \rho_{n_i + \Delta n_i}(\sqrt{m_{i,1}};-1) = \frac{\abs{m_{i,1}}^{n_i + \Delta n_i}}{(n_i + \Delta n_i)!} \e^{-\abs{m_{i,1}}} \\
	&\approx \frac{1}{\sqrt{2\pi m_{i,1}}}\e^{-(n_i + \Delta n_i - m_{i,1})^2/(2m_{i,1})} \numberthis \\
	&\pi \rho_{n_i}(\sqrt{m_{i,2}};-1) = \frac{\abs{m_{i,2}}^{n_i}}{n_i!} \e^{-\abs{m_{i,2}}} \\
	&\approx\frac{1}{\sqrt{2\pi m_{i,2}}}\e^{-(n_i - m_{i,2})^2/(2m_{i,2})} \numberthis
.\end{align*}
Inserting this in the definition of the Skellam distribution from Eq.~\eqref{eq:skellam_dist}, it takes the form
\begin{align*}\label{eq:skellam_approx_theta}
    &p_\text{s}(\Delta n_i; m_{i,1}, m_{i,2}) \\
	&=\pi^2 \sum_{n_i=\tilde{n}_i}^\infty \rho_{n_i + \Delta n_i}(\sqrt{m_{i,1}};-1)\rho_{n_i}(\sqrt{m_{i,2}};-1) \\
	&\approx \frac{1}{2\pi\sqrt{m_{i,1}m_{i,2}}} \e^{-m_{i,2}/2-(\Delta n_i-m_{i,1})^2/(2m_{i,1})} \\
	&\vartheta_3\left[\frac{1}{2\iu\pi}(2-\Delta n_i/m_{i,1}),-\frac{1}{\pi \iu}\frac{m_{i,1}+m_{i,2}}{2m_{i,1}m_{i,2}}\right] \\
	&= \frac{1}{\sqrt{2\pi(m_{i,1}+m_{i,2}})}\e^{-(m_{i,1}-m_{i,2} - \Delta n_i)^2/(2m_{i,1}+2m_{i,2})}  \\
	&\times \vartheta_3\left[\frac{(\Delta n_i-2m_{i,1})m_{i,2}}{m_{i,1}+m_{i,2}},\pi \iu\frac{2m_{i,1}m_{i,2}}{m_{i,1}+m_{i,2}}\right] \numberthis
,\end{align*}
where $\vartheta_3\left(z;\tau\right) = \sum_{n \in \Z} \e^{\pi \iu n^2 \tau + 2\pi n \iu z} = 1 + 2\sum_{n = 1}^\infty \e^{\pi \iu n^2 \tau}\cos(2 \pi nz)$ is the Jacobi theta function. In the last step of Eq.~\eqref{eq:skellam_approx_theta}, the identity $\vartheta_3\left(z;\tau\right) = \frac{1}{\sqrt{-\iu\tau}}\e^{-\pi \iu z^2/\tau} \vartheta_3\left(\frac{z}{\tau};\frac{-1}{\tau}\right)$ was used. This justifies the approximation $\vartheta_3\left(z;\tau\right) \approx 1$ of the theta function, because for strong probe amplitudes, the second argument of the theta function is $\tau = \pi \iu\frac{2m_{i,1}m_{i,2}}{m_{i,1}+m_{i,2}} \sim \iu\abs{\beta_{i}}^2$. 

\section{Post-measurement quasiprobability distribution}\label{a:post_meas}
To calculate the $(s_\text{X},s_\text{Y})$-parametrized quasiprobability distribution, with $s_\text{Q} < 2\mu^2/(1+A_Q)-1$, the anti-normally ordered characteristic function of the post-measurement state is needed:
\begin{align*}\label{eq:post_characteristic}
	&\chi^\prime(\xi;-1) = \tr(\hat{\rho}_\Omega^\prime \e^{-\conj{\xi}\hat{a}_\Omega} \e^{\xi \hat{a}_\Omega^\dagger}) = \frac{1}{\pi p(\{\Delta n_i\})} \int \rho(\gamma;1) \\
	&\times \int \e^{2\iu\Im[\xi\conj{(\gamma^\prime)}]} \sum_{\{n_i\}}\abs{\prescript{}{\Omega}{\bra{\gamma^\prime}}\hat{M}_{\{n_i,\Delta n_i\}}\ket{\gamma}_\Omega}^2 \dd\gamma^\prime\dd\gamma \numberthis
.\end{align*}
Expressing the Fock states again using the Glauber-Sudarshan distribution, as done in appendix \ref{a:count_prob}, the characteristic function is
\begin{align*}\label{eq:post_characteristic_2}
	&\chi^\prime(\xi;-1) = \frac{1}{\pi p(\{\Delta n_i\}} \int \rho(\gamma;1) \int \e^{2\iu\Im[\xi\conj{(\gamma^\prime)}]} \\
	&\times \sum_{\{n_i\}} \iint \abs{\prescript{}{\Omega}{\bra{\gamma^\prime}}\prescript{}{i,s}{\bra{\eta_{i,s}}}\prescript{}{i,z}{\bra{\eta_{i,z}}}\hat{U}\ket{0}_{i,z}\ket{0}_{i,s}\ket{\gamma}_\Omega}^2 \\
	&\times \rho_{n_i+\Delta n_i}(\eta_{i,s};1)\rho_{n_i}(\eta_{i,z};1) \dd \eta_{i,s} \dd \eta_{i,z} \dd\gamma^\prime\dd\gamma \numberthis
.\end{align*}
The matrix elements are already known from count-probability distribution and thus the problem can be solved by applying the same approximation of the Skellam distribution as before. With this approximation, the characteristic function is
\begin{align*}
	&\chi^\prime(\xi;-1) \approx \left(p^2(\{\Delta n_i\}) (1+A_X)(1+A_Y)\prod_i 2 \pi \abs{\beta_i}^2\right)^{-\frac{1}{2}} \\
	&\times \exp[-\sum_i \Delta n_i^2/(2\abs{\beta_i}^2)] \int \rho(\gamma;1) \exp\Bigg\{-\abs{\gamma}^2\\
	&+\left[\frac{\Re(\gamma)}{\mu} + \iu\Im(\xi) + \frac{\abs{\nu}}{\mu} \sum_{i \in I_\text{X}}\frac{\abs{\tilde{\alpha}_i}}{\abs{\beta_i}} \Delta n_i\right]^2\frac{\mu^2}{1+A_X} \\
	&+\left[\frac{\Im(\gamma)}{\mu} - \iu\Re(\xi) + \frac{\abs{\nu}}{\mu} \sum_{i \in I_\text{Y}}\frac{\abs{\tilde{\alpha}_i}}{\abs{\beta_i}} \Delta n_i\right]^2\frac{\mu^2}{1+A_Y} \Bigg\} \dd \gamma \numberthis
.\end{align*}
The $(s_\text{X},s_\text{Y})$-parametrized quasiprobability distribution can then be calculated from the anti-normally ordered characteristic function using the Fourier transformation:
\begin{align*}
	&\rho^\prime(z;s_\text{X},s_\text{Y}) \\
	&= \frac{1}{\pi^2} \int \e^{z\conj{\xi} - \conj{z}\xi} \e^{\frac{1+s_\text{Y}}{2}\Re^2(\xi) + \frac{1+s_\text{X}}{2}\Im^2(\xi)} \chi^\prime(\xi;-1) \dd \xi \\
	&\approx N^\prime(z;s_\text{X},s_\text{Y}) \rho(z^\prime(z);s_\text{X}^\prime,s_\text{Y}^\prime) \numberthis
.\end{align*}
This is also where the restrictions $s_\text{Q} < 2\mu^2/(1+A_Q)-1$ originate from, because only for those values do the integrals in the above equation converge.

\section{Approximation of the Fidelity}\label{ap:fidelity}
Since the coherent states $\ket{\alpha_\Omega}$ are over-complete $\int \rho_{\alpha_{\Omega}}(z;s_\text{X},s_\text{Y}) \dd \alpha_{\Omega} = 1$, the parameter distribution is just:
\begin{equation}
	p(\alpha_\Omega | \{\Delta n_i\}) = \rho_{\alpha_\Omega}(z;s_\text{X},s_\text{Y})
\end{equation}
and hence, the reconstructed state is
\begin{equation}
	\hat{\rho}_\text{rec} = \int \rho_{\alpha_\Omega}(z;s_\text{X},s_\text{Y}) \ket{\alpha_\Omega}_\Omega \prescript{}{\Omega}{\bra{\alpha_\Omega}} \dd \alpha_\Omega
.\end{equation}
Now we can calculate the fidelity between the initial and reconstructed state:
\begin{align*}
	&F(\hat{\rho}_\Omega,\hat{\rho}_\text{rec}) = \int \rho_{\alpha_\Omega}(z;s_\text{X},s_\text{Y}) \abs{\braket{\alpha}{\alpha_\Omega}}^2 \dd \alpha_\Omega\\
	&= 2\frac{\sqrt{(1-s_\text{X})(1-s_\text{Y})}}{\sqrt{(2/(1-s_\text{X})+1)(2/(1-s_\text{Y})+1)}} \\
	&\times \exp\Big(\Re^2(\frac{2}{1-s_\text{X}}z + \alpha_\Omega)/(\frac{2}{1-s_\text{X}}+1) \\
	&+ \Im^2(\frac{2}{1-s_\text{Y}}z + \alpha_\Omega)/(\frac{2}{1-s_\text{Y}}+1)  \\
	&- \frac{2}{1-s_\text{X}}\Re^2(z) - \frac{2}{1-s_\text{Y}}\Im^2(z) - \abs{\alpha_\Omega}^2\Big) \numberthis
.\end{align*}
To calculate the average fidelity between the initial and reconstructed state, the same approximation for the sum as in section~\ref{s:ensemle} is applied. Using this approximation, the average fidelity for a $\overline{\text{XY}}$-measurement, as well as for a $\overline{\text{XYXY}}$-measurements of a coherent state $\ket{\alpha_\Omega}_\Omega$ is
\begin{align*}
	&\ev{F(\hat{\rho}_\Omega,\hat{\rho}_\text{rec})} = \sum_{\Delta n_\text{X},\Delta n_\text{Y}} p(\Delta n_\text{X}, \Delta n_\text{Y})F(\hat{\rho}_\Omega,\hat{\rho}_\text{rec}) \\
	&\approx \pi \int \rho_{\alpha_\Omega}(z;s) F(\hat{\rho}_\Omega,\hat{\rho}_\text{rec}) \dd z \\
	&= \frac{1}{\pi} \frac{1}{s-2} = \left[2\coth^2(\abs{\zeta})+1\right]^{-1} \numberthis
\end{align*}
and can be seen in Fig.~\ref{fig:comparison_sim} (a) as a function of the squeezing $\abs{\zeta}$. The limit of this function for $\abs{\zeta} \to \infty$ is 1/3.\\
For a consecutive $\overline{\text{XY}} \rightarrow \overline{\text{XY}}$ measurements of a coherent state $\ket{\alpha_\Omega}_\Omega$ the parameter distribution is:
\begin{equation}
	p(\alpha | \{\Delta n_i\},\{\Delta n_i^\prime\} = \frac{\rho_{\alpha_\Omega}(z;s_\text{X},s_\text{Y})\rho_{\alpha_\Omega}(z^\prime;s_\text{X}^\prime,s_\text{Y}^\prime)}{\int \rho_{\alpha_\Omega}(z;s_\text{X},s_\text{Y})\rho_{\alpha_\Omega}(z^\prime;s_\text{X}^\prime,s_\text{Y}^\prime) \dd \alpha_\Omega}
,\end{equation}
with
\begin{align*}
	&\int \rho_{\alpha_\Omega}(z;s_\text{X},s_\text{Y})\rho_{\alpha_\Omega}(z^\prime;s_\text{X}^\prime,s_\text{Y}^\prime) \dd \alpha_\Omega \\
	&= \frac{1}{\pi}\frac{(1-s)(1-s^\prime)}{2/(1-s)+2/(1-s^\prime)}\exp\Bigg(\abs{\frac{2}{1-s}z + \frac{2}{1-s^\prime}z^\prime}^2 \\
	&\times \left(\frac{2}{1-s} + \frac{2}{1-s^\prime}\right)^{-1} - \frac{2\abs{z}^2}{1-s} - \frac{2\abs{z^\prime}^2}{1-s^\prime}\Bigg) \numberthis
.\end{align*}
The fidelity for a consecutive $\overline{\text{XY}} \rightarrow \overline{\text{XY}}$ measurements of a coherent state $\ket{\alpha_\Omega}_\Omega$ is therefore:
\begin{align*}
	&F(\hat{\rho}_\Omega,\hat{\rho}_\text{rec}) = \frac{\int \rho_{\alpha_\Omega}(z;s) \rho_{\alpha_\Omega}(z^\prime;s) \abs{\braket{\alpha}{\alpha_\Omega}}^2 \dd \alpha}{\int \rho_{\alpha_\Omega}(z;s_\text{X},s_\text{Y})\rho_{\alpha_\Omega}(z^\prime;s_\text{X}^\prime,s_\text{Y}^\prime) \dd \alpha_\Omega} \\
	&= \left(\frac{2}{1-s} + \frac{2}{1-s^\prime}\right)\left(\frac{2}{1-s} + \frac{2}{1-s^\prime} + 1\right)^{-1} \e^{\abs{\alpha_\Omega}^2} \\
	&\times \exp\Bigg(\abs{\frac{2}{1-s}z + \frac{2}{1-s^\prime}z}^2\left(\frac{2}{1-s} + \frac{2}{1-s^\prime}\right)^{-1} \\
	&+\abs{\frac{2}{1-s}z + \frac{2}{1-s^\prime}z + \alpha_\Omega}^2\left(\frac{2}{1-s} + \frac{2}{1-s^\prime} + 1\right)^{-1}\Bigg) \numberthis 
.\end{align*}
The average fidelity can then be obtained using the same approximation as for the single EOS-measurement ($z_1 = \frac{1}{\sqrt{2}\abs{\nu}\abs{\beta}}(\Delta n_\text{X} + \iu \Delta n_\text{Y})$, $z_1 = \frac{1}{\sqrt{2}\abs{\nu}\abs{\beta}}(\Delta n_\text{X}^\prime + \iu \Delta n_\text{Y}^\prime)$):
\begin{align*}
	&\ev{F(\hat{\rho}_\Omega,\hat{\rho}_\text{rec})} = \hspace{-1cm}\sum_{\Delta n_\text{X},\Delta n_\text{Y},\Delta n_\text{X}^\prime,\Delta n_\text{Y}^\prime} \hspace{-1cm} p_1(\Delta n_\text{X}, \Delta n_\text{Y})p_2(\Delta n_\text{X}^\prime, \Delta n_\text{Y}^\prime) \\
	&\times F(\hat{\rho}_\Omega,\hat{\rho}_\text{rec}) \approx \frac{4\pi^2\abs{\beta}^4\abs{\nu}^4}{4\pi\abs{\beta}^4\abs{\nu}^2}\frac{1}{\pi^2}\frac{\abs{\nu}^4}{\mu^4}\frac{1}{\abs{\nu}^2/\mu^2+1} \\
	&\times \iint \exp\Bigg(-\abs{\nu}^2(\abs{z_1}^2 + \abs{z_2}^2 - \abs{z_1+z_2}^2/2) \\
	&- \frac{\abs{\nu}^2}{\mu^2}\abs{\alpha_\Omega - (z_1+z_2)/\sqrt{2}}^2 - \frac{\abs{\nu}^2}{2\mu^2}\abs{z_1 + z_2}^2 - \abs{\alpha_\Omega}^2 \\
	&+ \abs{\alpha_\Omega - \frac{\abs{\nu}^2}{\mu^2}(z_1+z_2)/\sqrt{2}}^2(\abs{\nu}^2/\mu^2 + 1)^{-1} \Bigg) \dd z_1 \dd z_2 \\
	&=\frac{1}{\pi}\frac{\sinh^2(\abs{\zeta})}{2\cosh^6(\abs{\zeta})} \left[1+\sinh^2(\abs{\zeta})+\frac{1}{2}\sinh^4(\abs{\zeta})\right] \numberthis
.\end{align*}
See Fig.~\ref{fig:comparison_cons} for a plot as a function of the squeezing $\abs{\zeta}$. The limit of this function for $\abs{\zeta} \to \infty$ is 2/5. This is above the limit for the simultaneous measurement.

\bibliography{discrete_eos_final_EH_TG}

\end{document}